\documentclass[aps,prd,twocolumn,groupedaddress,nofootinbib]{revtex4}
\pdfoutput=1  % togliere il commento se le figure sono pdf, lasciarlo se sono eps
\usepackage{bm}
\usepackage{graphicx,color}
\usepackage{latexsym,amsmath,amssymb,graphicx,booktabs}
\usepackage{hyperref}
\usepackage{placeins}
\usepackage{bm}
\usepackage{subfigure}
\definecolor{MyBlue}{rgb}{0.15,0.15,0.70}
\definecolor{Dgreen}{rgb}{0,0.7,0.0}

\hypersetup{
colorlinks=true,
citecolor=MyBlue,
linkcolor=MyBlue,
urlcolor=MyBlue
}

\usepackage{amssymb}
\usepackage{amsmath}
\usepackage{amsfonts}
\usepackage{upgreek}
\usepackage{latexsym}
\usepackage{appendix}
\usepackage{eufrak}
\usepackage{dsfont}

\usepackage[export]{adjustbox}

\include{mydefs}

\newcommand\spart{\;\raise1.0pt\hbox{/}\hskip-6pt\partial}
\newcommand\spartb{\;\overline{\raise1.0pt\hbox{/}\hskip-6pt\partial}}

\newcommand{\be}{\begin{equation}}
\newcommand{\ee}{\end{equation}}
\newcommand{\bk}{{\bm k}}

\newcommand{\beqa}{\begin{eqnarray}}
\newcommand{\eeqa}{\end{eqnarray}}

\newcommand{\bx}{{\bm{x}}}
\newcommand{\bA}{{\bm{A}}}
\newcommand{\bee}{{\bm{e}}}

\newcommand{\bv}{{\bm{v}}}

\newcommand{\bp}{{\bm{p}}}

\newcommand{\bGamma}{{\bf{\Gamma}}}

\newcommand{\dd}{\text{d}}

\newcommand{\nn}{\nonumber}

\newcommand{\obs}{_{\rm O}}
\newcommand{\Gal}{_{\rm G}}
\newcommand{\Sou}{_{\rm S}}

\newcommand{\Ka}{_{\rm K}}
\newcommand{\bbB}{{\bf{ B}}}

\graphicspath{ {./Graphs/} }
\begin{document}
\title{Anisotropy of the astrophysical gravitational wave background:\\ analytic expression of the angular power spectrum and correlation with cosmological observations}
\author{Giulia Cusin}
\email{giulia.cusin@unige.ch}
\affiliation{D\'epartement de Physique Th\'eorique and Center for Astroparticle Physics, Universit\'e de Gen\`eve, 24 quai Ansermet, CH--1211 Gen\`eve 4, Switzerland}
\author{Cyril Pitrou}
\email{ pitrou@iap.fr}
\affiliation{Institut d'Astrophysique de Paris, CNRS UMR 7095, Universit\'e Pierre \& Marie Curie - Paris VI, 98 bis Bd Arago, 75014 Paris, France \\
           Sorbonne Universit\'es, Institut Lagrange de Paris, 98 bis, Bd Arago, 75014 Paris, France}
\author{Jean-Philippe Uzan}
\email{uzan@iap.fr}
\affiliation{Institut d'Astrophysique de Paris, CNRS UMR 7095, Universit\'e Pierre \& Marie Curie - Paris VI, 98 bis Bd Arago, 75014 Paris, France \\
           Sorbonne Universit\'es, Institut Lagrange de Paris, 98 bis, Bd Arago, 75014 Paris, France}
\vspace{1 em}
\date{\today}

\begin{abstract}
Unresolved sources of gravitational waves are at the origin of a stochastic gravitational wave background. While the computation of its mean density as a function of frequency in a homogeneous and isotropic universe is standard lore, the computation of its anisotropies requires to understand the coarse graining from local systems, to galactic scales and then to cosmology. An expression of the gravitational wave energy density valid in any general spacetime is derived. It is then specialized to a perturbed Friedmann-Lema\^{\i}tre spacetime in order to determine the angular power spectrum of this stochastic background as well as its correlation with other cosmological probes, such as the galaxy number counts and weak lensing. Our result for the angular power spectrum also provides an expression for the variance of the gravitational wave background.
\end{abstract}
\maketitle
%\tableofcontents

%%%%%%%%%%%%%%%%%%%%%%%%%%%%%%%%%%%%
\section{Introduction}\label{sec_intro}
%%%%%%%%%%%%%%%%%%%%%%%%%%%%%%%%%%%%

Astronomical observations include the detection of radiation from resolved objects as well as various stochastic backgrounds of radiations (electromagnetic, gravitational \dots), due to the superposition of the signals from all unresolved sources. The electromagnetic backgrounds include the cosmic microwave background with a black body spectrum at $2.725$~K and the extragalactic background, made up of all the light emitted by stars, galaxies, quasars etc. since their formation. The analysis of these backgrounds provides information on the dynamics of our universe and on the distribution and evolution of its large scale structure. Similarly, there exists a neutrino background, the observation of which seems to be out of reach, and a background of  gravity waves (GW). 

The recent detection by the Advanced Laser Interferometric Gravitational-wave Observatory (LIGO) of the gravitational wave sources GW150914~\cite{Abbott:2016blz} provided the first observation of the merging of a binary black hole (BBH) system. This first detection has been followed by other two observations of  similar systems, i.e.  GW151226 \cite{Abbott:2016nmj}  and GW170104 \cite{Abbott:2017vtc} and by the very recent observation of a black hole merging from both the LIGO and VIRGO detectors \cite{Abbott:2017oio}. 
 Following these observations, the rate and mass of coalescing binary black holes appear to be greater than many previous expectations. As a result, the stochastic background from unresolved compact binary coalescences is expected to be particularly loud. In \cite{TheLIGOScientific:2016dpb}  a search for the isotropic stochastic gravitational-wave background is performed using data from Advanced LIGO's first observing run. The dimensionless energy density of gravitational waves is constrained to be $\Omega_{GW}<1.7 \times 10^{-7}$ with 95\% confidence, assuming a flat energy density spectrum in the most sensitive part of the LIGO band (20-86 Hz). This is a factor of $\sim$33 times more sensitive than previous measurements ~\cite{Aasi:2014zwg,TheLIGOScientific:2016dpb}          
and his improves bounds on the stochastic background obtained from the analysis of big-bang nucleosynthesis~\cite{Maggiore:1999vm, Allen:1996vm},  and of the cosmic microwave background~\cite{Smith:2006nka,Henrot-Versille:2014jua} at 100~Hz. At low frequencies, Pulsar Timing Array (see below) gives a bound  $\Omega_{\rm GW}<1.3\times10^{-9}$ for $f=2.8 \times 10^{-9}$~Hz~\cite{Shannon:2013wma}.

The possibility of measuring and mapping the gravitational wave background is discussed in Refs.~\cite{Allen:1996gp, Cornish:2001hg, Mitra:2007mc, Thrane:2009fp, Romano:2015uma, Romano:2016dpx} and a discussion of the different methods  which can be used by LIGO and LISA (Laser Interferometer Space Antenna) to reconstruct an angular resolved map of the sky are presented in Ref.~\cite{TheLIGOScientific:2016xzw}. An analogous discussion for  Pulsar Timing Arrays  can be found in Refs. \cite{Mingarelli:2013dsa,Taylor:2013esa, Gair:2014rwa}. 

Stochastic GW backgrounds are made up of the superposition of astrophysical signals from unresolved sources. The different backgrounds contribute at different frequencies and have distinct statistical properties, making it potentially possible to distinguish them~\cite{Regimbau:2011rp}. In the standard cosmological model~\cite{PeterUzan2005}, the existence of a primordial GW background from the amplification of vacuum quantum fluctuations is a generic prediction of any inflationary phase. Gravitational waves may also be produced at the end of inflation during the reheating phase, see e.g. Ref.~\cite{Dufaux:2007pt} for an analytic and numerical study of the process. More speculative sources of an early GW background include pre big-bang modes, cosmic strings~\cite{Vilenkin:1981bx, Hogan:1984is, Vachaspati:1984gt, Caldwell:1991jj, Kuroyanagi:2016ugi}, first order phase transitions in the early universe ~\cite{Caprini:2009fx, Caprini:2015zlo}, magnetic fields \cite{Caprini:2001nb}; see Ref.~\cite{Binetruy:2012ze} for a review on those topics and  Refs.~\cite{Maggiore:1999vm, Buonanno:2014aza} for general reviews. In addition, an astrophysical contribution  (AGWB) results from the superposition of a large number of unresolved sources since the beginning of stellar activity. The nature of the AGWB may differ from its cosmological counterpart, expected to be stationary, unpolarized, statistically Gaussian and isotropic, by analogy with the cosmic microwave background. Many different sources may contribute to the AGWB, including black holes and neutron star mergers~\cite{TheLIGOScientific:2016wyq, Regimbau:2016ike, Mandic:2016lcn, Dvorkin:2016okx, Nakazato:2016nkj, Dvorkin:2016wac, Evangelista:2014oba}, supermassive black holes~\cite{Kelley:2017lek},  exploding supernovae (SNe)~\cite{Crocker:2015taa}, neutron stars~\cite{Surace:2015ppq, Talukder:2014eba, Lasky:2013jfa},  and stellar core collapse~\cite{Crocker:2017agi}, population III binaries~\cite{Kowalska:2012ba}.

The observational landscape is also growing and covers large bands of frequencies; see e.g. Ref.~\cite{Moore:2014lga} for a review\footnote{The associated code {\tt http://rhcole.com/apps/GWplotter/} allows one to generate plots of  noise curves for many detectors and associated target sources.}.  \textcolor{black}{At extremely low frequencies $\sim 10^{-16}$ Hz bounds come mainly from the analysis of CMB B-modes} while at  low frequency of order $10^{-10}-10^{-6}$~Hz, there are pulsar timing arrays such as the radio telescope Parks Pulsar Timing Array\footnote{{\tt http://www.atnf.csiro.au/research/pulsar/ppta/}} (PPTA), the Large European Array for Pulsar Timing\footnote{{\tt http://www.leap.eu.org}} (LEPTA) and the future International Pulsar Timing Array\footnote{{\tt http://www.ipta4gw.org}} (IPTA). At low frequencies (typically $10^{-6}-10^{0}$~Hz) detection relies on space-borne detectors, such as Laser Interferometer Space Antenna\footnote{{ www.lisamission.org}} (LISA) and the evolved Laser Interferometer Space Antenna\footnote{{\tt https://www.elisascience.org}} (eLISA) launched in 2016. High frequency (typically $10^{0}-10^{5}$~Hz) observations rely on ground-based detectors, such as  LIGO and its advanced configuration (aLIGO), VIRGO\footnote{{\tt https://www.ego-gw.it/public/about/whatIs.aspx}}, the Einstein Telescope\footnote{{\tt http://www.et-gw.eu}} (ET) \textcolor{black}{or its american counterpart Cosmic Explorer (CE)} \cite{Evans:2016mbw}. This spectrum covers most of the theoretical predictions.

So far, constraints have been set on the total energy density of GW background integrated over the sky, as a function of frequency, $\Omega_{\rm GW}(\nu)$, even though some constraints on the anisotropy have been obtained by PTA~\cite{Taylor2015,Sesana2008},  (see below for definition) and by the advanced LIGO first observing run \cite{TheLIGOScientific:2016dpb}. On the theoretical side, the energy density of GW has been modeled and parameterized under the assumption that both our universe and the distribution of sources are homogeneous and isotropic, see e.g. Refs.~\cite{Regimbau:2011rp,Dvorkin:2016okx}. These assumptions can be relaxed in order to take into account that astrophysical sources are located in cosmic structures that indeed have a distribution that can be computed in a given cosmological model. Therefore the flux of energy from all astrophysical sources (resolved and unresolved) is not constant across the sky and depends on the direction of observation. The goal of our work is to present an analytic framework to describe and compute the anisotropy in the observed energy density of the AGWB, taking into account  the presence of inhomogeneities in the matter distribution and geometry in our observed universe. 

\begin{figure*}
\includegraphics[scale=0.30]{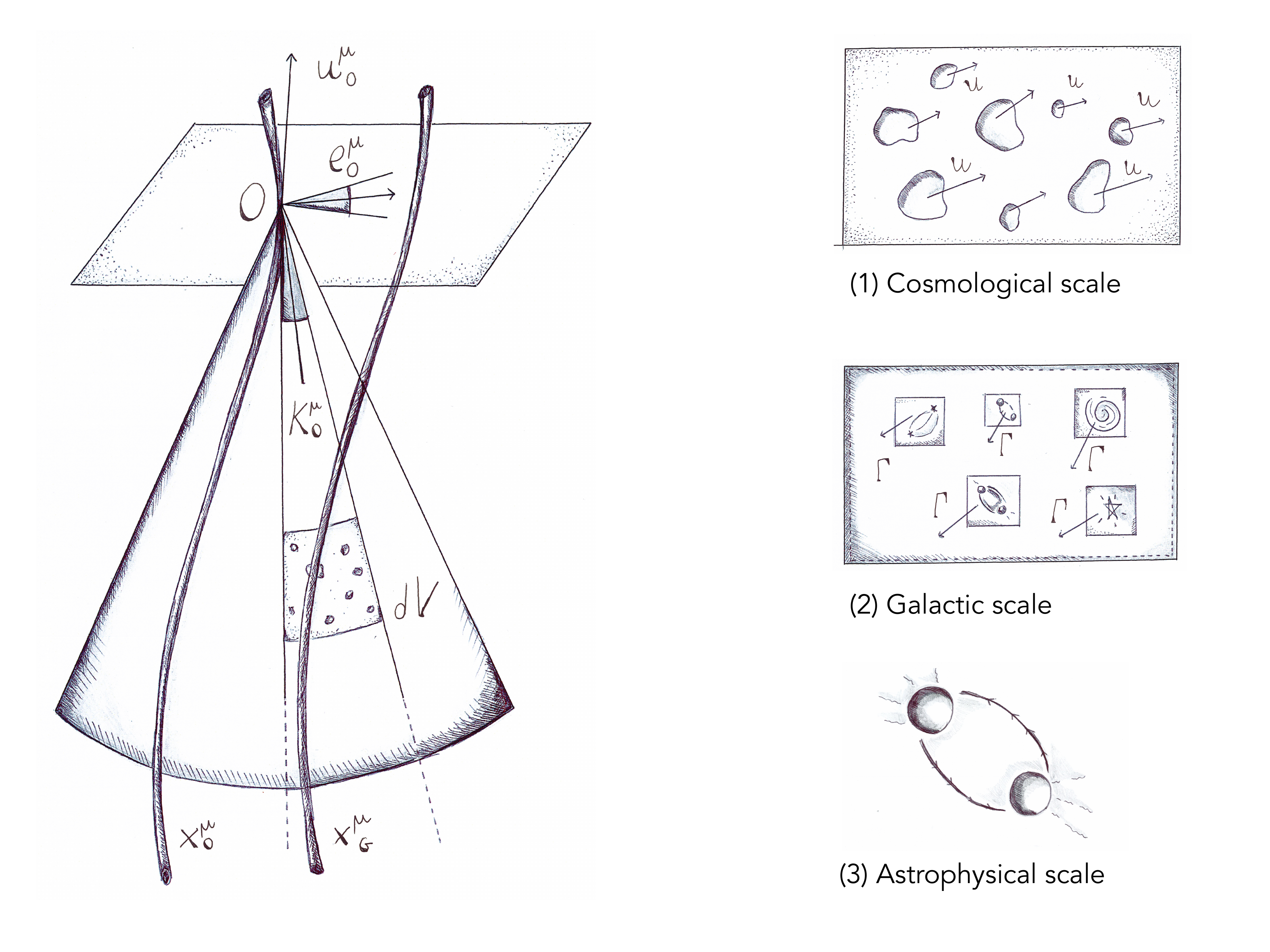}
\caption{\label{figure}
Schematic representation of our approach.
Left: Spacetime representation. The observer detects a signal in the direction $e_{\obs}^{\mu}$, with solid angle $\dd^2\Omega_{\obs}$ around this direction. This defines a bundle of null geodesics along its past lightcone. The physical volume element $\dd V$ is a 3-dimensional volume element, defined as the intersection of a 4-dimensional volume element with the observer past lightcone. Right: The three scales in our problem: cosmological, galactic and astrophysical.  }
\end{figure*}

To that goal, we need to distinguish different scales in our problem, represented in Fig. \ref{figure}. The first and largest scale is the {\em cosmological scale} of the observer for whom galaxies are point-like GW sources. The total flux of GW he/she receives is the sum of the fluxes of all individual galaxies. The flux from a given galaxy is related to the galaxy effective (GW) luminosity. This is a physical quantity defined in the galaxy rest frame, that takes into account contributions from the various sources inside the galaxy ({\em galactic scale}). Each of these sources has a peculiar motion with respect to the galaxy rest frame, inducing an extra dispersion in the signal. The luminosity of an astrophysical source in the galaxy rest frame can in turn be expressed in terms of local quantities characterizing the source in its rest frame ({\em local astrophysical scale}). In such a way, one can express cosmological observables characterizing GW at the observer scale in terms of quantities defined on local scales of single GW sources.  In other words, we shall provide a final parameterization for the observed GW flux, which inherits two main contributions: the first has a  \emph{cosmological} origin, i.e. it is related to the specific cosmology we are looking at (i.e. geometry of the universe, distribution of the large scale structure). The second comes from \emph{local physics} and encodes information on the specific processes of GW emission. This latter contribution depends on the details of the emission processes and on the intrinsic properties of GW sources inside a galaxy (e.g. rates and densities of the different kinds of processes).  

It follows that the anisotropies of the AGWB encode information about (1) the different mechanisms of GW production,  (2) the astrophysical distribution of GW sources in galaxies, (3) galaxy formation and distribution of the large scale structures of the universe and (4) the spacetime geometry along the line of sight. The question is which of these effects dominates on different frequency bands. In particular, if local effects are dominant, the AGWB will teach us mostly about the distribution of GW sources  while if it is the cosmological contribution to dominate, we shall have a new cosmological observable robust to local physics unknown.  The answer to this question will be discussed in our companion article~\cite{Cusinprep} , where we also analyze the shape and amplitude of the AGWB angular power spectra for different frequencies together with  the correlation with other cosmological probes.  Our theoretical framework can be straightforwardly generalized to discuss different cosmological scenarios: for example, if black hole binary systems are of stellar origin then the AGWB is expected to be correlated with the galaxy distribution while if dark matter is related to a 20-100 solar mass primordial BH component~\cite{Carr:2016drx,Bird:2016dcv,Clesse:2016ajp} the correlation between AGWB and the galaxy distribution will be smaller, see e.g. Ref.~\cite{Raccanelli:2016cud}, where the possibility of distinguishing these models is discussed. The correlation of resolved BH binaries with
galaxy clustering and weak lensing is also discussed in Ref. \cite{Namikawa:2016edr}, which in particular
discusses the possibility for 2nd generation GW detectors to measure such correlations.

Our approach relies on integrating all sources. Alternatively, one can also derive an effective source term for a Boltzmann approach to the AGWB, as tried in Ref.~\cite{Contaldi:2016koz}. For resolved sources, our analysis provides a modellisation of the GW luminosity distance~\cite{Namikawa:2015prh,Bertacca:2017vod,Olmez:2011cg}, see  also Ref. \cite{Oguri:2016dgk} for the cross-correlation of these GW standard sirens to galaxy clustering. On the other side, the total energy density of the AGWB depends only on the ratio between the luminosity distance $D_L$ and the angular diameter distance $D_A$. This ratio is given by $(1+z)^2$ because of the  reciprocity relation, which holds in any spacetime geometry \cite{Uzan:2004my, Etherington1933}.

This study is organized as follows. In Section~\ref{general} we introduce the line of sight approach, we define the various scales of the problem and derive a parameterization for the energy density of the AGWB  valid in a  generic cosmology. We also describe the coarse graining necessary to link observable quantities to local quantities characterizing the processes of GW emission. In Section~\ref{cosmological} we specialize this result to the case of a perturbed Friedmann-Lema\^{\i}tre-Robertson-Walker (FLRW) cosmology. This provides all the tools to compute the angular power spectrum of the AGWB energy density in Section~\ref{angular}. To conclude, in Sections~\ref{lensing} and~\ref{counts}, we compute
the correlators with other cosmological probes, namely weak-lensing by the large scale structure and galaxy number counts. The applicability of this formalism is discussed in Section~\ref{conclusions}, where we pave the way to the  further numerical investigations presented in our companion article~\cite{Cusinprep}.\\

{\bf{Notations}}: The cosmic scale factor is normalized to unity at the time of observation, $a_{\obs}=1$, and we use units defined such that $c=\hbar=k_{B}=1$. A dot corresponds to a derivative with respect to conformal time, i.e. $\dot X=\partial_{\eta}X$. 

%%%%%%%%%%%%%%%%%%%%%%%%%%%%%%%%%%%%
\section{General framework}\label{general}
%%%%%%%%%%%%%%%%%%%%%%%%%%%%%%%%%%%%

\textcolor{black}{The spectrum of the AGWB is characterized by the dimensionless density parameter
\be\label{Ogw}
\Omega_{\rm GW}(\nu_{\obs})\equiv \frac{\nu_{\obs}}{\rho_c}\frac{\dd\rho_{\rm GW}(\nu_{\obs})}{\dd\nu_{\obs}}\,, 
\ee
where $\rho_c\equiv 3H_0^2/(8\pi G)$ is the critical density, $\nu_{\obs}$ is the frequency as measured by the observer, and $\rho_{\rm GW}$ is the AGWB energy density~\cite{Maggiore:1900zz}. The corresponding spectrum $\dd\rho_{\rm GW}/\dd\nu_{\obs}$ is related to the directional dependent spectrum, whose general derivation is the goal of this section, through
\be\label{drhodnu}
\frac{\dd\rho_{\rm GW}(\nu_{\obs})}{\dd\nu_{\obs}} = \int \frac{\dd^3\rho_{\rm GW}(\nu_{\obs},e_{\obs})}{\dd\nu_{\obs}\dd^2 \Omega_{\obs}} \dd^2 \Omega_{\obs}\,,
\ee
where $e_{\obs}$ is a spatial 3-vector defining the direction of observation (see Eq.~(\ref{e.def-e}) below for definition).
} 

\subsection{Line of sight approach to GW propagation}

We work in the eikonal limit in which the propagation of GW is described in a similar way as light in the geometrical optic approximation~\cite{Maggiore:1900zz,DeruelleBook,1961RSPSA.264..309S,Isaacson:1967zz}. The spacetime metric can be decomposed as
\be\label{gh}
(g_{\mu\nu})_{\text{tot}}=g_{\mu\nu}+h_{\mu\nu}\,,
\ee
where $h_{\mu\nu}$ is the GW metric perturbation and we refer to $g_{\mu\nu}$ as \emph{general background metric} (e.g. in cosmology it will contain the background FLRW metric as well as cosmological perturbations).  

Einstein equations imply that GW follow null geodesics~\cite{Maggiore:1900zz,DeruelleBook}, $x^\mu(\lambda)$ where  $\lambda$ is an affine parameter along the geodesic. Its tangent vector 
\be\label{kmudxmudl}
k^{\mu}=\frac{\dd x^{\mu}}{\dd\lambda}
\ee 
is a null vector ($k^{\mu}k_{\mu}=0$) that satisfies the geodesic equation
\be
\frac{D k^{\mu}}{D\lambda}\equiv k^{\nu}\nabla_{\nu}k^{\mu}=0\,,
\ee
where $\nabla_{\nu}$ denotes the covariant derivative associated to the metric $g_{\mu\nu}$.

Consider an observer with 4-velocity $u^{\mu}$ ($u_{\mu}u^{\mu}=-1$). At any time, his worldline is the origin of the observer past lightcone containing all observed GW rays, see Figure \ref{figure}. The 4-velocity $u^\mu$ defines a preferred spatial section and the spatial direction of GW propagation, defined as the opposite of the direction of propagation of the signal converging to the observer. It is spanned by the purely spatial unit vector $e^{\mu}$\,,
\be\label{e.def-e}
e^{\mu}u_{\mu}=0\,,\hspace{1 em}e^{\mu}e_{\mu}=1\,,
\ee
which provides the 3+1 decomposition of the wave 4-vector \footnote{In other terms, introducing a space and a time projectors, $\mathcal{S}_{\mu\nu}=g_{\mu\nu}+u_{\mu}u_{\nu}$ and $\mathcal{T}_{\mu\nu}=-u_{\mu}u_{\nu}$ respectively, we have $e^{\mu}\equiv -\mathcal{S}^{\mu}_{\nu}k^{\nu}/E$.}
\be\label{directionobs}
k^{\mu}=E \left(u^{\mu}-e^{\mu}\right)\,,
\ee
where $E=2 \pi \nu\equiv -u_{\mu}k^{\mu}$ is the cyclic frequency of the GW signal in the observer's rest frame. We define the spatial projection of the wave 4-vector as 
\be\label{p}
p^{\mu}\equiv \left(g^{\mu\nu}+u^{\mu}u^{\nu}\right) k_{\nu}=-E e^{\mu}\,. 
\ee
The redshift $z_{\Gal}$ is defined from the ratio between the emitted frequency $\nu_{\Gal}$ in the source's rest frame and the observed frequency in the observer's rest frame $\nu_{\obs}$ as
\be\label{redshift}
1+z_{\Gal}\equiv \frac{\nu_{\Gal}}{\nu_{\obs}}=\frac{u_{\Gal}^{\mu}\,k_{\mu}(\lambda_{\Gal})}{u_{\obs}^{\mu}\,k_{\mu}(\lambda_{\obs})}\,,
\ee
where $u^{\mu}_{\Gal}$ is the 4-velocity of the source and $u^{\mu}_{\obs}$ is the 4-velocity of the observer. The source $G$ \footnote{As discussed in the introduction, we identify sources with galaxies containing astrophysical sources of GW.} located at a redshift $z_{\Gal}$ is emitting GW with a given frequency spectrum. From the definition~(\ref{redshift}), it follows that the frequency measured in $O$, $\nu_{\obs}$, is related to the frequency at the emission, $\nu_{\Gal}$ by 
\be\label{nus}
\nu_{\Gal}=(1+z_{\Gal})\nu_{\obs}\,.
\ee
Let us conclude by stressing that $x^{\mu}(\lambda)$ is the worldline of a graviton which intersects the worldline of the observer at the time of observation.  Then, by definition,
\be\label{11}
x^{\mu}(\lambda_{\obs})=x_{\obs}^{\mu}\,,\hspace{1 em} \left. \frac{\dd x^{\mu}(\lambda)}{\dd\lambda}\right|_{\lambda=\lambda_{\obs}}=E_{\obs}(u_{\obs}^{\mu}-e_{\obs}^{\mu})\,.
\ee
Therefore, $x^{\mu}(\lambda)$ is a function of the direction of observation and of 4-position of the observer, i.e. $x^{\mu}(\lambda)=x^{\mu}(\lambda, e^{\mu}_{\obs}, x^{\mu}_{\obs})$. In the following, to make the notation compact, the dependence on $e^{\mu}_{\obs}$ and $x^{\mu}_{\obs}$ will be understood.

\subsection{Parameterization of the observed GW radiation}

As discussed in the introduction, providing a general parameterization for the GW energy flux from unresolved astrophysical sources requires to consider 3 scales.
\begin{enumerate}
\item \emph{cosmological scale}. The observer measures a GW flux  in a direction $e_{\obs}$  and  solid angle $\Omega_{\obs}$. The angular resolution of the observer is such that we assume galaxies to be point-like sources emitting GW and comoving with the cosmic flow.
\item \emph{galactic scale.} A galaxy is described by a set of parameters such as its mass, mean metallicity, \dots We associate to each galaxy an effective luminosity encoding the global effects  of all the GW sources contained into it. These sources have a velocity $\bGamma$ in the galaxy rest frame. Therefore,  to get the effective luminosity of each type of astrophysical sources, we need to sum the luminosities of all single sources of that type, averaged over their velocity distribution function in the galaxy rest frame. 
\item \emph{astrophysical scale.} This is the local scale of single GW sources. For any source (binary inspiralling, binary merging, supernova, \dots) one can study the astrophysical process of GW production and calculate its energy spectrum, function of local parameters characterizing the source. 
\end{enumerate}

We denote as $i-$type source a specific source of GW. Each type of source is fully characterized by specific values for its local parameters $\theta^{(i)}$ (e.g. orbital parameters, masses \dots) and by the parameters characterizing  its host galaxy (mass $M_{\Gal}$, metallicity $Z_{\Gal}$, \dots), which we will indicate collectively as $\theta_{\Gal}$. These parameters are in full generality time-dependent functions. We consider $\theta_{\Gal}$ and $\theta^{(i)}$ evaluated at the time at which the signal received by the observer $O$ at the time of observation $t_{\obs}$,  has been emitted. 

Let us now turn to the definitions of the physical quantities on each scale and on their connections.

\subsubsection{ Cosmological scale}\label{cosmologyscale}

Since the intrinsic angular resolution of present observations does not allow to resolve galaxies, we describe them as point-like sources emitting GW and comoving with the cosmic flow. Each galaxy is characterized by its   \emph{effective luminosity}, which depends on its parameters $\theta_{\Gal}$, and that needs to be expressed in terms of the properties of the sources contained in the galaxy. 

Consider a galaxy  observed at a redshift $z_{\Gal}$ in direction $e_{\obs}$. The total flux of energy received by the observer $O$ in this direction is related to the absolute \emph{effective}  luminosity of the galaxy (i.e. the power radiated by the galaxy, in the galaxy rest frame) by the following relation
\be\label{lum}
\Phi(z_{\Gal}, e_{\obs}, \theta_{\Gal})\equiv \frac{L_{\Gal}(\theta_{\Gal})}{4\pi D_{\rm L}^2(z_{\Gal}, e_{\obs})} \,,
\ee
where $D_{\rm L}(z_{\Gal}, e_{\obs})$ is the luminosity distance of the galaxy. The angular diameter distance  is defined as
\be
D_{\rm A}^2\equiv \frac{A_{\Gal}}{\Omega_{\obs}}\,,
\ee
where $A_{\Gal}$ is the physical area of the galaxy and $\Omega_{\obs}\ll 1$ is its apparent observed angular size.  Luminosity and angular diameter distances are related by the reciprocity relation (see e.g. \cite{Etherington1933, Uzan:2004my, Bassett:2003vu, Ellis:2013cu} for a test of this relation)
\be\label{reciprocity}
D_{\rm L}=(1+z_{\Gal})^2 D_{\rm A}\,. 
\ee

Let us introduce the effective frequency spectrum of GW  of the galaxy,  $\mathcal{L}_{\Gal}(\nu_{\Gal}, \theta_{\Gal})$, normalized in such a way that
\be
\int_0^{\infty} \mathcal{L}_{\Gal}(\nu_{\Gal}, \theta_{\Gal})\,\dd\nu_{\Gal}=L_{\Gal}(\theta_{\Gal})\,.
\ee
Making use of  Eq.~(\ref{nus}),  we find  
\begin{equation}
L_{\Gal}(\theta_{\Gal})=(1+z_{\Gal})\int  \mathcal{L}_{\Gal}[\nu_{\obs}(1+z_{\Gal}), \theta_{\Gal}] \dd\nu_{\obs}\,.
\end{equation}
We can then define the flux measured by the observer in the frequency range $(\nu_{\obs}, \nu_{\obs}+\dd\nu_{\obs})$. Using the definition (\ref{lum}), it follows that 
\be\label{specflux}
\Phi(z_{\Gal}, e_{\obs}, \nu_{\obs}, \theta_{\Gal})d\nu_{\obs}\equiv \frac{(1+z_{\Gal})}{4\pi D_{\rm L}^2(z_{\Gal}, e_{\obs})}\mathcal{L}_{\Gal}(\nu_{\Gal}, \theta_{\Gal})  \dd\nu_{\obs}\,.
\ee
This quantity is called \emph{specific flux} of GW radiation.

\subsubsection{Galactic scales}

We now need to relate the former quantities to those associated to the astrophysical GW sources inside each given galaxy. In the galaxy rest frame,  one can introduce a local system of coordinates (different from the cosmological coordinate system). In this system, a GW source characterized by $\theta^{(i)}$ has a velocity $\bGamma(\theta^{(i)}, \theta_{\Gal})$, that also depends on the parameters of the host galaxy (in particular the velocity dispersion depends on the total mass).

We assume the velocities to have a distribution function $f(\bGamma, \theta_{\Gal})$ where the explicit dependence on $\theta_{\Gal}$ reminds that different galaxies have in principle associated different distribution functions. The function $f$ is normalized as
\be
\int \dd^3\bGamma\,f(\bGamma, \theta_{\Gal})=1\,,
\ee
and we assume that each source type is globally at rest with respect to the galaxy,  i.e.
\be\label{0}
\int \dd^3\bGamma\,\bGamma f(\bGamma, \theta_{\Gal})=0\,.
\ee
To describe the collective emission of all the sources of a galaxy, one should first perform a Lorentz transformation to shift from the reference frame of the source to the one of the galaxy (described e.g. by the normal vector to the galactic plane) to describe the emission of a single source in the galactic rest frame. Then, one shall perform a rotation to describe the emission in galactic coordinates.  These transformations are described in Appendix~\ref{Lorentz}. However, these effects are linear in $\bGamma$ at lowest order. When averaging over all the sources of a given galaxy, this lowest order contribution will cancel thanks to Eq.~(\ref{0}) and higher contributions will be negligible since we expect $|\bGamma|/c\ll1$.

We denote as $\bA^{(i)}_{\Gal}$ the axis characterizing the orientation of the source in this galactic frame. In the source rest frame, we expect the GW emission to be rotationally invariant around the axis of the source $\bA^{(i)}_{\Sou}$, i.e. the spectrum depends on the  direction $\bee_{\Sou}^{(i)}$ only through $\mu_{\Sou}^{(i)}=\cos \alpha^{(i)}_{\Sou}=\bee_{\Sou}\cdot \bA_{\Sou}^{(i)}$. As explained in detail in the Appendix, 
%the galactic frame is related to the source local frame by a rotation (aligning the $z$ axis with the axis of the galaxy) and a boost with velocity $\bGamma$.   As a consequence,
 the luminosity  emitted by a given GW source in the galaxy rest frame, $L_{\Gal}^{(i)}$, is a function of the velocity of the source and of its orientation in the galaxy rest frame, $\bA_{\Gal}^{(i)}$,  and of the line of sight in the galaxy frame $\bee_{\Gal}$. Explicitly, the effective luminosity per unit solid angle 
$\dd^2\Omega_{\Gal}$ (defined in the galaxy rest frame), is
\be\label{above2}
\frac{\dd^2 L_{\Gal}^{(i)}}{\dd^2\Omega_{\Gal}}\equiv \frac{\dd^2 L_{\Gal}^{(i)}}{\dd^2\Omega_{\Gal}}(\bGamma, \theta^{(i)},  \bee_{\Gal}, \bA_{\Gal}^{(i)}, \theta_{\Gal})\,,
\ee
where the subscript $G$ indicates again that we are defining quantities in the galaxy rest frame. 

In full generality, the computation of the total flux in the galaxy rest frame requires both the information on the orientation  $\bA_{\Gal}^{(i)}$ of each system and its velocity $\bGamma(\theta^{(i)}, \theta_{\Gal})$ in the galaxy rest frame. In the following, we assume that $\bA_{\Gal}^{(i)}$  and $\bGamma(\theta^{(i)}, \theta_{\Gal})$ are uncorrelated random variables on which we average. Physically, the average on $\bA_{\Gal}^{(i)}$ corresponds to an average over the possible orientations of a given source. 
%We then integrate over the direction $\Omega_{\Gal}$. 
This integration  is justified by the fact that we do not have access to the information on the distribution of sources inside a galaxy. All this leads to the fact that the effective luminosity of the galaxy is isotropic in its rest frame, justifying our notation $L_{\Gal}(\theta_{\Gal})$\footnote{Note that if this is not the case, one would also need to know the orientation of each galaxy with respect to the line of sight!}. Under this hypothesis, each galaxy is a point source radiating isotropically in its rest frame.

The GW sources can be split in two categories, depending on the duration of the emission process with respect to the characteristic time of evolution of the system. We consider 
\begin{itemize}
\item \emph{inspiralling binary sources}, for which the evolution of the orbital parameters is slow compared to galactic time scales so that the emission of GW can be averaged on many periods. To compute the luminosity associated to inspiralling systems, we need to compute the number of systems and the radiated power;
\item \emph{merging binary sources and exploding SNe}, for which the duration of the emission is short compared to the typical galactic time scale. To compute the luminosity associated to these systems, we need to  determine the rate of events and the total radiated energy. 
\end{itemize}
We shall thus decompose the total effective luminosity of a galaxy as
\be\label{IM}
L_{\Gal}(\theta_{\Gal})=L^I_{\Gal}(\theta_{\Gal})+L^M_{\Gal}(\theta_{\Gal})+L^{SN}_{\Gal}(\theta_{\Gal})\,,
\ee
respectively indicating the contributions from inspiralling binaries (I),  mergers (M)  and supernovae (SN). As we shall see, the last two contributions have formally the same structure. However, to keep the notation clear, we write them separately. 

The first contribution can be decomposed as
\begin{eqnarray}
L_{\Gal}^I(\theta_{\Gal})&\equiv &\int_0^{\infty}\dd\nu_{\Gal} \mathcal{L}_{\Gal}^I(\nu_{\Gal}, \theta_{\Gal})\,,\label{up}\\
&=&\sum_{(i)}^I\int \dd\theta^{(i)}\, \mathcal{N}^{(i)}(\theta^{(i)}, \theta_{\Gal})\int \dd^3\bGamma\,f(\bGamma, \theta_{\Gal})\nn \\
&&\hspace{- 4 em}\times \int \frac{\dd^2\Omega_{\Gal}}{4\pi}\, \dd^2\bA_{\Gal}^{(i)} \frac{\dd^2 L_{\Gal}^{(i)}}{\dd^2\Omega_{\Gal}}(\bGamma, \theta^{(i)}, \bee_{\Gal},  \bA_{\Gal}^{(i)},  \theta_{\Gal})\,,\nn\\\label{up00}
\end{eqnarray}
where the sum  runs over different types of inspiralling binary systems (e.g. back hole-black hole, neutron star-neutron star \dots). In Eq. (\ref{up00}), $\mathcal{N}^{(i)}(\theta^{(i)}, \theta_{\Gal})$ is the number of sources in the inspiralling phase with parameters $\theta^{(i)}$ in a galaxy with parameters $\theta_{\Gal}$ and $L_{\Gal}^{(i)}$ is the effective luminosity of a single source inside the galaxy. Note that in full generality we should have considered $\mathcal{N}^{(i)}(\theta^{(i)}, \theta_{\Gal},\bA_{\Gal}^{(i)})$ but we assumed the distribution of the axis to be isotropic. It follows that the dependence on the direction of emission drops out in the integration over $\dd^2\Omega_{\Gal}$  so that the final result, $L_{\Gal}^I(\theta_{\Gal})$, reduces to a monopole for the whole galaxy.

To grasp the physical meaning of the single source effective luminosity  (per unit of solid angle $\dd^2\Omega_{\Gal}$) and to understand how this can be related to the power radiated by the source, we express it as
\begin{align}\label{chiara}
&\frac{\dd^2L_{\Gal}^{(i)}}{\dd^2\Omega_{\Gal}}(\bGamma, \theta^{(i)}, \bee_{\Gal},  \bA_{\Gal}^{(i)},  \theta_{\Gal})\equiv \nn\\
&\quad \equiv \int \dd\nu_{\Gal}\,\frac{\dd^2 \mathcal{L}_{\Gal}^{(i)}}{\dd^2\Omega_{\Gal}}(\nu_{\Gal}, \bGamma, \theta^{(i)},  \bee_{\Gal}, \bA_{\Gal}^{(i)}, \theta_{\Gal})\\
&\quad =\int \dd\nu_{\Gal}\,\frac{\dd^4 E^{(i)}_{\Gal}}{\dd^2\Omega_{\Gal} \dd t_{\Gal} \dd\nu_{\Gal}}(\nu_{\Gal}, \bGamma, \theta^{(i)}, \bee_{\Gal}, \bA_{\Gal}^{(i)}, \theta_{\Gal})\,,
\end{align}
where the power per unit of solid angle $\dd E_{\Gal}^{(i)}/(\dd t_{\Gal}\dd\nu_{\Gal}\dd^2\Omega_{\Gal})$ in the galaxy rest frame is related to the emitted power per unit solid angle (in the source local frame) through a rotation and Lorentz transformation. Inserting Eq.~(\ref{chiara}) in Eq.~(\ref{up00}), we can read the following parameterization for the effective luminosity per unit of effective frequency 
\begin{eqnarray}\label{LI}
\mathcal{L}_{\Gal}^I\left(\nu_{\Gal}, \theta_{\Gal}\right)&\equiv& \sum_{(i)}^I\int \dd\theta^{(i)} \mathcal{N}^{(i)}(\theta^{(i)}, \theta_{\Gal})\int \dd^3\bGamma f\left(\bGamma, \theta_{\Gal}\right)\nn\\
&& \hspace{-6 em} \times 
\int \frac{\dd^2\Omega_{\Gal}}{4\pi}\ \dd^2\bA_{\Gal}^{(i)} \frac{\dd^4 E^{(i)}_{\Gal}}{\dd^2\Omega_{\Gal}\dd t_{\Gal} d\nu_{\Gal}}(\nu_{\Gal}, \bGamma, \theta^{(i)}, \bee_{\Gal},  \bA_{\Gal}^{(i)}, \theta_{\Gal})\,.\nn\\
\end{eqnarray}

The second contribution to Eq.~(\ref{IM}) from mergers can be decomposed as
\begin{eqnarray}\label{up2}
L_{\Gal}^M(\theta_{\Gal})&\equiv& \int_0^{\infty}\dd\nu_{\Gal} \mathcal{L}_{\Gal}^M(\nu_{\Gal}, \theta_{\Gal})\,,\\
&=&\sum_{(i)}^M\int \dd\theta^{(i)}\, \frac{\dd\mathcal{N}^{(i)}}{\dd t_{\Gal}}(\theta^{(i)}, \theta_{\Gal})\int \dd^3\bGamma\,f(\bGamma, \theta_{\Gal})\nn\\
&&\hspace{-5 em} \times \int \dd\nu_{\Gal}\frac{\dd^2\Omega_{\Gal}}{4\pi} \dd^2\bA_{\Gal}^{(i)}\frac{\dd^3 E_{\Gal}^{(i)}}{\dd^2\Omega_{\Gal}\dd\nu_{\Gal}} (\nu_{\Gal}, \bGamma,  \theta^{(i)},  \bee_{\Gal}, \bA_{\Gal}^{(i)}, \theta_{\Gal})\,,\label{up22}\nn\\
\end{eqnarray}
where the sum runs over all types of mergers (i.e. back hole-black hole, neutron star-neutron star \dots) and $\dd\mathcal{N}^{(i)}/\dd t_{\Gal}$ is the merging rate of systems of type-$i$ in the galaxy rest frame. $\dd E_{\Gal}^{(i)}/\dd\nu_{\Gal}$ is the energy spectrum  per unit solid angle (in the galaxy rest frame) from a source of type-$i$ with parameters $\theta^{(i)}$ and velocity $\bGamma$ in the galaxy rest frame. The latter is related to the emitted spectrum in the source local frame through a Lorentz transformation.\footnote{The rate of mergers can be further parametrized as $$ \frac{\dd\mathcal{N}^{(i)}}{\dd t_{\Gal}}(\theta^{(i)}, \theta_{\Gal})={\rm SFR} \times\mathcal{N}^{(i)}\left(\theta^{(i)}, \theta_{\Gal}\right)\,,$$  where  the star formation rate (SFR) function depends on the details of local physics. } 
Comparing Eq. (\ref{up2}) and Eq. (\ref{up22}), we can read the following expression for the effective luminosity associated to merging objects, per unit of effective frequency
\begin{align}\label{LM}
&\mathcal{L}_{\Gal}^M(\nu_{\Gal}, \theta_{\Gal})=\nn\\
&\sum_{(i)}^M\int \dd\theta^{(i)}\, \frac{\dd\mathcal{N}^{(i)}}{\dd t_{\Gal}}(\theta^{(i)}, \theta_{\Gal}) \int \dd^3\bGamma\,f(\bGamma, \theta_{\Gal})\nn\\
&\times \int \frac{\dd^2\Omega_{\Gal}}{4\pi} \dd^2\bA_{\Gal}^{(i)}\frac{\dd^3 E_{\Gal}^{(i)}}{\dd^2\Omega_{\Gal} \dd\nu_{\Gal}}\left(\nu_{\Gal}, \bGamma,  \theta^{(i)}, \bee_{\Gal}, \bA_{\Gal}^{(i)},  \theta_{\Gal} \right)\,.\nn\\
\end{align}

The decomposition of the third contribution to Eq. (\ref{IM}) from SNe follows the same steps described above for the case of merging binary systems and the final result is the analogous of Eq. (\ref{LM}), with the sum running over the different types of SNe and with the quantity $\dd\mathcal{N}^{(i)}/\dd t_{\Gal}$ denoting the explosion rate of SNe of type-$i$ in the galaxy rest frame.

Equations (\ref{LI}) for inspirals and (\ref{LM}) for merger and supernovae, are the main results of this paragraph. 

\subsubsection{Astrophysical scale}

For each galaxy and each type of astrophysical sources, one can calculate the GW spectrum or power emitted by the source in its local frame as a function of the source local parameters. Local quantities in the source frame, that we denote with a subscript $S$,  are related to quantities in the galaxy rest frame through a Lorentz transformation, once the velocity of the source with respect to the galaxy rest frame is known. This is detailed in Appendix \ref{Lorentz}. The effect of the velocity of the center of mass of a binary on its wavefront is discussed in Ref.~\cite{Bonvin:2016qxr}.

At linear order in the velocity $\bGamma$, the effects of the peculiar motion of a source in its host galaxy can be neglected on average. Explicitly, the effective luminosity of a galaxy per unit of effective frequency, Eqs. (\ref{LI}) and  (\ref{LM})  can be rewritten in terms of local quantities at the source since at linear order in $\bGamma$,  $\nu_{\Gal}=\nu_{\Sou}$, $E_{\Gal}=E_{\Sou}$, and $t_{\Gal}=t_{\Sou}$, and the directional dependence reduces to a dependence in $\mu_{\Gal}^{(i)} \equiv \bee_{\Gal}\cdot \bA_{\Gal}^{(i)}$ if in the source rest frame the dependence on directions is only through $\mu_{\Sou}^{(i)}$ (see Appendix~\ref{Lorentz}).

\subsubsection{Summary: from astrophysical to cosmological scale}

We can now gather all these pieces to derive the expression of the flux of energy received by the observer $O$ in the direction $e_{\obs}$, per unit of observed frequency in terms of quantities defined at the small scales  (galactic and astrophysical). The definition ~(\ref{specflux}) gives the specific flux
\begin{align}\label{int2}
\Phi&(z_{\Gal}, e_{\obs}, \nu_{\obs}, \theta_{\Gal})\dd\nu_{\obs}=\frac{(1+z_{\Gal})\mathcal{L}_{G}(\nu_{\Gal}, \theta_{\Gal})}{4\pi D_{\rm L}^2(z_{\Gal}, e_{\obs})}  \dd\nu_{\obs}\,,
\end{align}
where the total effective luminosity per unit of effective frequency $\mathcal{L}_{\Gal}$ can be decomposed as the sum of contributions from inspiralling binaries, mergers and SNe  as
\be\label{LGG}
\mathcal{L}_{G}\equiv \mathcal{L}^{I}_{\Gal}+\mathcal{L}^{M}_{\Gal}+\mathcal{L}^{SN}_{\Gal}\,,
\ee
where $\mathcal{L}^{I, M, SN}_{\Gal}$, are respectively defined in Eqs. (\ref{LI}) and (\ref{LM}), and can be expressed in terms of local quantities defined in the local frame of the GW sources inside the galaxy.

\subsection{Final results}\label{final}

Equation (\ref{int2}) provides a general expression for the specific flux (i.e. flux per unit of observed frequency) of the gravitational waves received by an observer $O$ in a direction $e_{\obs}$  from a  galaxy at redshift $z_{\Gal}$, assumed to be a point-like source and isotropically radiating in its rest frame.

In practice, because of  limited spatial resolution, an instrument responds to the flux per solid angle, i.e. the \emph{intensity} of GW from the source.\footnote{What is actually measured in an image is the \emph{specific intensity} $\mathcal{I}$, i.e. the intensity in a specific frequency range and solid angle around the direction of observation.}  Our goal is to derive the total energy density of GW received in a solid angle $\dd^2\Omega_{\obs}$ around the direction of observation $e_{\obs}$ per unit of observed frequency $\nu_{\obs}$. This requires to add the specific fluxes received from the various galaxies contained in this solid angle.

 Formally, we need to consider the null geodesic $x^{\mu}(\lambda)$ passing through $O$ in the direction $e_{\obs}$, i.e. satisfying  the conditions (\ref{11}). At any event corresponding to a given value of $\lambda$ of the affine parameter, the local physical 3-volume element is
\be\label{covV}
\dd^3 V\equiv \sqrt{-g}\,\epsilon_{\mu\nu\alpha\beta}\,u^{\mu}\dd x^{\nu}\dd x^{\alpha}\dd x^{\beta}\,. 
\ee
The total intensity of GW observed in the direction $e_{\obs}$ is then easily obtained by summing the contributions from all the galaxies along the line of sight. Explicitly,  if we want the contribution between $\lambda$ and $\lambda+\dd\lambda$ we need to multiply the number of galaxies in $\dd^3 V$ by the specific flux received from a single galaxy with parameters $\theta_{\Gal}$, integrating  over the range of possible values for $\theta_{\Gal}$. One gets the final expression
 \begin{widetext}
 \be\label{master}
 \frac{\dd^3\rho_{\rm GW}}{\dd\nu_{\obs}\dd^2\Omega_{\obs}}(\nu_{\obs}, e_{\obs})=\int \dd\lambda \int \dd\theta_{\Gal}\, \Phi\left[x^{\mu}(\lambda), \nu_{\obs}, \theta_{\Gal}\right] \,\frac{\dd^3\mathcal{N}_{\Gal}}{\dd\lambda\,\dd^2\Omega_{\obs}}\left[x^{\mu}(\lambda), \theta_{\Gal}\right]\,,
 \ee
\end{widetext}
where $\Phi[x^{\mu}(\lambda), \nu_{\obs}, \theta_{\Gal}]$ is the specific flux received at frequency $\nu_{\obs}$ in direction $e_{\obs}$ from a galaxy located at $x^{\mu}(\lambda)$, defined in Eq. (\ref{int2}) and the direction dependent spectrum on the left hand side is related to the GW energy density in Eq. (\ref{drhodnu}). $\dd^3 \mathcal{N}_{\Gal}(x^{\mu}(\lambda), \theta_{\Gal})$ represents the number of galaxies with parameters $\theta_{\Gal}$ contained in the physical volume $\dd^3 V$ defined in Eq. (\ref{covV}), seen by the observer $O$ under the solid angle $\dd^2 \Omega_{\obs}$. We introduce $n_{\Gal}\left[x^{\mu}(\lambda), \theta_{\Gal}\right]$, the physical density of galaxies with parameters $\theta_{\Gal}$, defined by
\be\label{NG}
\dd^3\mathcal{N}_{\Gal}\left[x^{\mu}(\lambda), \theta_{\Gal}\right]\equiv n_{\Gal}\left[x^{\mu}(\lambda), \theta_{\Gal}\right] \dd^3 V\left[x^{\mu}(\lambda)\right]\,.
\ee
To simplify our final result, it is useful to rewrite Eq. (\ref{covV}) expressing the physical volume element as 
\be\label{Volume}
\dd^3 V\left[x^{\mu}(\lambda)\right]=\dd^2\Omega_{\obs} D_{\rm A}^2(\lambda) \sqrt{p_{\mu}(\lambda) p^{\mu}(\lambda)} \dd\lambda \,,
\ee
where we used the fact that $\dd^3 V$ is the volume with cross-section $D_{\rm A}^2$ and depth $\sqrt{p_{\mu}(\lambda) p^{\mu}(\lambda)} \dd\lambda =\textcolor{black}{-(u_\mu k^\mu)}\dd\lambda$ along the line of sight, $p^{\mu}$ being defined in Eq. (\ref{p}). 

Substituting Eqs.~(\ref{NG}-\ref{Volume}) in Eq.~(\ref{master}) and using the expression~(\ref{int2}) for the specific flux, we get
 \begin{widetext}
 \be\label{mastermaster}
 \frac{\dd^3\rho_{\rm GW}}{\dd\nu_{\obs}\dd^2\Omega_{\obs}}(\nu_{\obs}, e_{\obs})=\frac{1}{4\pi}\int \dd\lambda \int \dd\theta_{\Gal}\, \frac{\sqrt{p_{\mu}(\lambda) p^{\mu}(\lambda)}}{\left[1+z_{\Gal}(\lambda)\right]^3}\,n_{\Gal}\left[x^{\mu}(\lambda), \theta_{\Gal}\right] \mathcal{L}_{\Gal}(\nu_{\Gal}, \theta_{\Gal})\,,
 \ee
\end{widetext}
where we have used the reciprocity relation (\ref{reciprocity}). In this equation, the total effective luminosity per unit of effective frequency,  $\mathcal{L}_{\Gal}$ is defined in Eq. (\ref{LGG}) as a sum of contributions from inspiralling binaries, mergers and  SNe. It depends on the local physics inside the galaxies while the others factors in the integral depend on the cosmology. 

In order to compare our expression with standard results in the literature, we split the GW energy density as
 \begin{align}\label{mm}
& \frac{\dd^3\rho_{\rm GW}}{\dd\nu_{\obs}\dd^2\Omega_{\obs}}(\nu_{\obs}, e_{\obs})\equiv \nn\\
 &\qquad\qquad\frac{\dd^3\rho^I_{\rm GW}}{\dd\nu_{\obs}\dd^2\Omega_{\obs}}+ \frac{\dd^3\rho^M_{\rm GW}}{\dd\nu_{\obs}\dd^2\Omega_{\obs}}+ \frac{\dd^3\rho^{SN}_{\rm GW}}{\dd\nu_{\obs}\dd^2\Omega_{\obs}}\,,
  \end{align}
 respectively for inspiralling binaries, mergers and SNe. The explicit expression of each one of these terms can be straightforwardly derived comparing (\ref{mm}) to Eq. (\ref{mastermaster}) with $\mathcal{L}_{G}\equiv \mathcal{L}^{I}_{\Gal}+\mathcal{L}^{M}_{\Gal}+\mathcal{L}^{SN}_{\Gal}$. 

We conclude observing that the expression (\ref{mastermaster}) as well as the definition (\ref{mm}) are completely general in the sense that they do not assume any specific form for the metric. They can be specialized to any cosmology.

\section{Cosmological framework}\label{cosmological}

We now consider the standard cosmological framework~\cite{PeterUzan2005} in which the universe is modeled by a FLRW universe with Euclidean spatial sections and with scalar perturbations.  In Newtonian gauge, the metric $g_{\mu\nu}$ is given by 
\be\label{FL}
\dd s^2=a^2\left[-(1+2\psi)\dd\eta^2+(1-2\phi)\delta_{ij}\dd x^i\dd x^j\right]\,,
\ee
where the metric of the constant time hypersurfaces is
\be
\delta_{ij}\dd x^i\dd x^j=\dd \chi^2+\chi^2\left(\dd\theta^2+\sin^2\theta\dd\phi^2\right)\,,
\ee
in terms of the comoving radial distance $\chi$. The two Bardeen potentials are decomposed as
\begin{equation}
 \psi = \Psi+\Pi,\qquad
 \phi = \Psi-\Pi.
\end{equation}
In the standard $\Lambda$CDM model, the matter content at late time is dominated by cold dark matter (CDM), described by a pressureless fluid, and by the cosmological constant. It follows that the Bardeen potentials,  $\phi$ and $\psi$, are equal, so that $\psi=\phi=\Psi$ and $\Pi=0$. 

We assume that the galaxies  are all comoving with the cosmic flow.\footnote{The velocity of galaxies is not biased $v(z\,,\bee)=v_{CDM}(z\,,\bee)$.}  To first order in perturbations, the four velocity of the cosmic fluid is given by
\be\label{peculiar}
u^{\mu}\equiv \frac{1}{a}(1-\psi\,, v^i)\equiv \bar u^\mu + \delta u^\mu\,,
\ee
where $v^i$ is the peculiar velocity field. From the matter conservation equation, the galaxy peculiar velocity can be related to the gravitational potential through the Euler equation.
%\footnote{The velocity of galaxies is not biased $v(z\,,\bee)=v_{CDM}(z\,,\bee)$.}\textcolor{red}{This footnote is not displayed}

\subsection{Strictly spatially homogeneous and isotropic  universe}

We start by considering an unperturbed FLRW universe (i.e. $\phi=\psi=0$ in Eq.~(\ref{FL})). Homogeneity and isotropy imply that  $\partial_i$ is the Killing vector associated to the invariance under spatial translations. Therefore, $g(\partial_i, k)=k_i$ is a constant of geodesic motion. Moreover, since $k$ is a null vector,  $E^2=g^{ij} k_i k_j$. From the geodesic equation, we get 
\be\label{kb}
\bar{k}^{\mu}(\lambda)=\frac{1}{a^2(\lambda)}\left(1, -\bee_{\obs}\right)\,,
\ee
where $\bee_{\obs}$ denotes the spatial direction of observation. From now on it will be denoted simply by $\bee$ and we denote with an overbar background quantities. It follows
\be
\bar{E}(\lambda)=\frac{1}{a(\lambda)}\,.
\ee
Further integrating Eq. (\ref{kb}), we find for the background geodesic
\begin{align}
\bar{x}^0(\lambda)-\bar{x}^0(\lambda_{\obs})&=\eta-\eta_{\obs}\equiv\Delta\eta\,,\\
\bar{x}^i(\lambda)-\bar{x}^i(\lambda_{\obs})&=e^i\Delta\eta\,.
\end{align}
The spatial direction of propagation of GW is defined in Eq. (\ref{p}) and now reads
\be\label{pp}
\bar{p}^{\mu}(\lambda)=\frac{1}{a^2(\lambda)}\left(0, -\bee\right)\,,
\ee
and the physical volume defined in Eq. (\ref{Volume}) reduces to
\be
\dd^3 V=\frac{1}{(1+\bar{z})^3}\frac{\chi^2(\bar{z})}{H(\bar{z})}\dd^2\Omega_{\obs} \dd\bar{z}\,
\ee
so that Eq.~(\ref{mastermaster}) reduces to
\begin{widetext}
\be\label{Irr}
\frac{\dd^3\rho_{\rm GW}}{\dd\nu_{\obs}\dd^2\Omega_{\obs}}(\nu_{\obs})=\frac{1}{4\pi H_0}\int \dd\bar{z}\frac{1}{E(\bar{z})}\frac{1}{(1+\bar{z})^4}\int \dd\theta_{\Gal}\,\bar{n}_{\Gal}(\bar{z}, \theta_{\Gal}) \mathcal{L}_{\Gal}(\nu_{\Gal}, \theta_{\Gal})\,.
\ee
\end{widetext}
In Eq. (\ref{Irr}), $\bar{n}_{\Gal}(\bar{z}, \theta_{\Gal})$ is the background (homogeneous and isotropic) density of galaxies. We have used that in a FLRW universe $D_{\rm L}(\bar z)=(1+\bar z)\chi(\bar z)$ and we have defined $E(\bar{z})\equiv H(\bar{z})/H_0$. As expected, this expression does not depend  on the direction of observation. When specialized to merging binaries (i.e. setting $\mathcal{L}_{\Gal}^{I, SN}=0$ inside $\mathcal{L}_{\Gal}$), Eq.~(\ref{Irr}) \textcolor{black}{once replaced in Eqs. (\ref{drhodnu}) and (\ref{Ogw})} is in agreement with the standard literature, e.g. with Eq.~(3.18) of  Ref.~\cite{Dvorkin:2016okx}~\footnote{In Ref.~\cite{Dvorkin:2016okx} the peculiar motion of sources  inside galaxies is not taken into account. Therefore, to compare our result with the one of \cite{Dvorkin:2016okx},  we need to set $f(\bGamma, \theta_{\Gal})=\delta(\bGamma)$ in Eq. (\ref{Irr}). As a consequence,  in our result we need to identify the frequency $\nu_{\Gal}$ with the frequency  $\nu_{\Sou}$. This explains the difference between our result (\ref{Irr}) and the one of Ref.~\cite{Dvorkin:2016okx} in the expression of the  local astrophysical contribution (i.e. inside $\mathcal{L}_{\Gal}$ in our expression). Moreover, in Ref.~\cite{Dvorkin:2016okx}, $n_{\Gal}$ is the comoving density of galaxy and not the physical one and the merging rate is defined with respect to the observer time and not the time at the source. This explains the additional factor of $(1+\bar{z})^{-4}$ in our expression.}.

\subsection{Perturbed FLRW universe}

We now consider the effect of the scalar perturbations in  the metric (\ref{FL}). Our goal is to obtain an expression for the GW background energy density  at first order in the metric perturbations.  Let us consider a graviton whose null worldline is intersecting the worldline of the observer at time $\eta_{\obs}$. We decompose it as
\be
x^{\mu}(\lambda)=\bar{x}^{\mu}(\lambda)+\xi^{\mu}(\lambda)\,,
\ee
where $\bar{x}^{\mu}$ is the null geodesic of the FLRW background spacetime and $\xi^{\mu}$ represents the deviation from the background geodesic. To solve the geodesic equation we use that geodesics are the same in conformally related metrics. We introduce a metric $\tilde{g}_{\mu\nu}$ such that $\dd s^2=a^2 \tilde{\dd s}^2$ and
\be
k^{\mu}=\frac{1}{a^2}\tilde{k}^{\mu}\,,\quad \lambda=a^2\tilde{\lambda}\,.
\ee
The null geodesic of the metric $\tilde{g}_{\mu\nu}$ reads
\be\label{deviation}
\frac{\dd\delta \tilde{k}^{\mu}}{\dd\tilde{\lambda}}\equiv \frac{\dd^2\tilde{\xi}^{\mu}}{\dd\tilde{\lambda}^2}=-\delta\tilde{\Gamma}^{\mu}_{\alpha\beta}\bar{\tilde{k}}^{\alpha}\bar{\tilde{k}}^{\beta}\,,
\ee
where an overbar indicates background quantities and $\delta\tilde{\Gamma}_{\alpha\beta}^{\mu}$ denotes the Levi-Civita connection of the $\tilde{g}_{\mu\nu}$ metric at first order in scalar perturbations. Solving Eq. (\ref{deviation}) and going back to the original metric $g_{\mu\nu}$, gives
\begin{align}
&\delta k^0(\lambda)=-\frac{2}{a^2}\left[\Psi+\Pi\right]_{\lambda_{\obs}}^\lambda+\frac{2}{a^2}\int_{\eta_{\obs}}^{\eta}\dd\eta' \dot{\Psi}\,,\label{k0}\\
&\delta k^i(\lambda)=\frac{2}{a^2}(\Pi-\Pi_{\obs}-\Psi)\,e^i-\frac{2}{a^2}\int_{\eta_{\obs}}^{\eta}\dd\eta'\, \partial_i\Psi \,,
\end{align}
where we have imposed initial condition at $\lambda=\lambda_{\obs}$ in such a way that the null-geodesic condition $k_{\mu}k^{\mu}=0$ is satisfied up to first order in perturbations. From Eq.~(\ref{kmudxmudl}), $\dd x^i/\dd \eta = k^i/k^0$, hence we obtain
\be
\xi^i=- 2  \int _{\eta_{\obs}}^{\eta} \dd\eta' \left[e^i\Psi+(\eta-\eta')(\delta_i^j-e_i e^j)\partial_j \Psi\right]\,,\label{xii}
\ee
where we have set $\xi^{i}(\eta_{\obs})=0$. 

The redshift of a graviton emitted from a galaxy at spacetime position $G$ and observed at $O$ is defined in Eq. (\ref{redshift}). Since gravitons follow null geodesics, we find
\be\label{expz}
1+z_{\Gal}=\frac{1}{a_{\Gal}}\left[1+(\Psi +\Pi-\bv\cdot \bee)|_{\Gal}^{\obs}+2 \int_{\eta_{\obs}}^{\eta_{\Gal}}\dd\eta \,\dot{\Psi}\right]\,
\ee
which we decompose as
\be
z_{\Gal}\equiv \bar{z}+\delta z\,,
\ee
where the first order correction  $\delta z$ is due to the metric perturbations and peculiar velocities. 

We now need to specialize Eq. (\ref{mastermaster})  to the perturbed metric (\ref{FL}) and expand the various functions in perturbations in order to derive the expression of the energy density received up to first order in perturbations. The density of galaxies $n(x^{\mu}, \theta_{\Gal})$ can be expanded as
\be\label{expnG}
n_{\Gal}(x^{\mu})=\bar{n}_{\Gal}(\bar{x}^{\mu})+\xi^{\mu}\nabla_{\mu}\bar{n}_{\Gal}(\bar{x}^{\mu})+\delta n_{\Gal}(\bar{x}^{\mu})\,,
\ee
where we have omitted the dependence on $\theta_{\Gal}$. The first term is the background contribution while the other two are first order corrections. Note that since $\bar{n}_{\Gal}$ is a background quantity, it does not depend on spatial coordinates, so that $\xi^{\mu}\nabla_{\mu}\bar{n}_{\Gal}=0$. We define
\be
\frac{\delta n_{\Gal}(\bar{x}^{\mu}, \theta_{\Gal})}{\bar{n}(\bar{x}^{\mu})}\equiv b(\eta)\delta_{\rm CDM}(\bar{x}^{\mu})\,,
\ee
where in full generality the bias $b$  depends on time coordinate; $\delta_{\rm CDM}$ denotes the CDM density contrast. Similarly, the physical volume in  $x^{\mu}(\lambda)$ can be expanded as 
\be
\dd^3 V(x^{\mu})=\dd^3\bar{V}(\bar{x}^{\mu})+\xi^{\mu}\nabla_{\mu}\dd^3\bar{V}(\bar{x}^{\mu})+\delta \dd^3 V(\bar{x}^{\mu})\,.
\ee
However, as explained in Section \ref{final}, by using the expression of the physical volume in terms of the observed solid angle and the angular diameter distance, Eq. (\ref{Volume}), the final expression for the total flux received, Eq. (\ref{mastermaster}), depends on the angular diameter distance only through the ratio $D_A/D_L$, given by $(1+z)^{-2}$ by the reciprocity relation (\ref{reciprocity}). We thus need to expand only the direction of propagation $p^{\mu}$ defined in Eq. (\ref{p}) up to first order in perturbations.  We get $p^{\mu}\equiv \bar{p}^{\mu}+\delta p^{\mu}$ with
$\bar p^\mu$ given in Eq. (\ref{pp}) and 
\begin{align}
\delta p^{0}&=-\frac{1}{a^2} \bv\cdot \bee\,,\\
\delta p^{i}&= - \frac{2}{a^2}\left(e^i (\Psi-\Pi+\Pi_{\obs})+\int _{\eta_{\obs}}^{\eta} d\eta' \partial_i \Psi +\frac{v^i}{2}\right)\,.
\end{align}
After simplifications, it can be shown that 
\begin{align}\label{expo}
&\sqrt{p^{\mu}p^{\nu}g_{\mu\nu}}=\frac{1}{a}+ \\
&+\frac{1}{a}\left[\Psi-\Pi+2\Pi_{\obs}+\bee\cdot \bv+2\int_{\eta_{\obs}}^{\eta} \dd \eta' \bee\cdot \nabla \Psi\right]\,.\nn
\end{align}
This quantity can be related to the ratio between proper time $\tau$ of the observer and conformal time $\eta$ as 
\begin{align}
\frac{\dd\tau}{\dd\eta}&\equiv \sqrt{S^{\mu}_{\alpha}S^{\nu}_{\beta}\frac{\dd x^{\alpha}}{\dd\eta}\frac{\dd x^{\beta}}{\dd\eta}g_{\mu\nu}}=  \sqrt{p^{\mu}p^{\nu}g_{\mu\nu}}\left(\frac{\dd\lambda}{\dd\eta}\right)\,, \nn\\
&=a\left[1+\Psi+\Pi+\bv\cdot \bee\right]\,.
\end{align}

Now,  we need to substitute Eqs.~(\ref{expz}),~(\ref{expnG}) and~(\ref{expo}) for the redshift, galaxy number density and for the norm of the direction of propagation of GW in Eq.~(\ref{mastermaster}) and keep terms up to first order in scalar perturbations. For redshift and galaxy density, we get, respectively 
\begin{widetext}
\begin{eqnarray}
1+z_{\Gal}&=&\frac{1}{a}\left[1+\Psi_{\obs}-\Psi+\Pi_{\obs}-\Pi-\bv_{\obs}\cdot \bee+ \bv\cdot \bee+2\int_{\eta_{\obs}}^{\eta}\dd\eta' \dot{\Psi}\right]\,,\label{3}\\
n_{\Gal}(x^{\mu})&=&\bar{n}_{\Gal}(\eta)+\delta n_{\Gal}\,. \label{2}
\end{eqnarray}
\end{widetext}
As a cross check,  one can compute $\sqrt{p^{\mu}p^{\nu}g_{\mu\nu}}\equiv -(u^{\mu}k_{\mu})=-(1+z_{\Gal}) (u^{\mu}k_{\mu})_{\obs}$ with $(1+z_{\Gal})$ given by eq. (\ref{3}) and verify that it reduces to the expression (\ref{expo}) for the norm of $p^{\mu}$. 
Gathering  Eqs.~(\ref{expo}),~(\ref{2}) and~(\ref{3}),  we get
\begin{widetext}
 \begin{equation} \label{product2}
\frac{\sqrt{p_{\mu}(\lambda) p^{\mu}(\lambda)}}{(1+z_{\Gal}(\lambda))^3}  n_{\Gal}[x^{\mu}(\lambda), \theta_{\Gal}]=\
a^2 \bar{n}_{\Gal}\left[1+b\delta_{\rm CDM}+2\Psi-\Psi_{\obs}+\textcolor{black}{2}\Pi-\Pi_{\obs}-2\bee\cdot \nabla v+3\bee\cdot \nabla v_{\obs} -4\int_{\eta_{\obs}}^{\eta}\dd\eta' \dot{\Psi}\right]\,.
\end{equation}
\end{widetext}
Then, inserting this result in Eq. (\ref{mastermaster})  for the total energy density of GW  and using the conformal time as integration variable, we get
\begin{widetext}
 \begin{eqnarray}\label{Tania2}
\frac{\dd^3\rho_{\rm GW}}{\dd\nu_{\obs}\dd^2\Omega_{\obs}}(\nu_{\obs}, \bee)&=&\frac{1}{4\pi}\int \dd\eta \,a^4 \int \dd\theta_{\Gal} \,\bar{n}_{\Gal}\,\mathcal{L}_{\Gal}(\nu_{\Gal}, \theta_{\Gal})\times\nn\\
&& \times \left[1+b\delta_{\rm CDM}+4\Psi+\textcolor{black}{4}\Pi-3\Psi_{\obs}-3\Pi_{\obs}-2\bee\cdot \nabla v+3\bee\cdot \nabla v_{\obs}-6\int_{\eta_{\obs}}^{\eta}\dd\eta' \dot{\Psi}\right]\,,
\end{eqnarray}
\end{widetext}
where we have substituted $\bv=\nabla v$ since we are considering only the contribution of scalar perturbations. The integration runs along the unperturbed geodesic, parameterized as $\bx=\bx_{\obs}+\bee(\eta_{\obs}-\eta)$.

We observe that in Eq. (\ref{Tania2}) the effective luminosity $\mathcal{L}{\Gal}(\nu_{\Gal}, \theta_{\Gal})$ is not a background quantity since $\nu_{\Gal}=(1+z_{\Gal}) \nu_{\obs}$ where the observed redshift $z_{\Gal}$ is given by Eq. (\ref{expz}) up to first order in perturbations. We write
\be
\nu_{\Gal}=\bar{\nu}_{\Gal}+\delta\nu_{\Gal}\,,
\ee
with
\be
\bar{\nu}_{\Gal}=(1+\bar{z})\nu_{\obs}\,,\quad \delta\nu_{\Gal}=\nu_{\obs} \delta z\,,
\ee
where $\bar{z}$ and $\delta z$ are defined in Eq. (\ref{expz}). We can therefore expand the effective luminosity in Eq. (\ref{Tania2}) around the background value of the effective frequency $\nu_{\Gal}$ as
\be
\mathcal{L}_{\Gal}(\nu_{\Gal}, \theta_{\Gal})=\mathcal{L}_{\Gal}(\bar{\nu}_{\Gal}, \theta_{\Gal})+\frac{\partial \mathcal{L}_{\Gal}}{\partial \nu_{\Gal}}(\nu_{\Gal}, \theta_{\Gal})\Big|_{\bar{\nu}_{\Gal}}\delta\nu_{\Gal}\,.
\ee
We plug this expansion in Eq. (\ref{Tania2}) and we split the result into a homogeneous and isotropic background contribution and a first order contribution as
\begin{eqnarray}\label{ss}
&&\frac{\dd^3 \rho_{\rm GW}}{\dd\nu_{\obs}\dd^2\Omega_{\obs}}(\eta_{\obs}, \bx_{\obs}, \bee, \nu_{\obs})=\\
&&\qquad  =\frac{\dd^3 {\bar\rho_{\rm GW}}}{\dd\nu_{\obs}\dd^2\Omega_{\obs}}(\eta_{\obs},\nu_{\obs})+\mathcal{E}(\eta_{\obs}, \bx_{\obs}, \bee,  \nu_{\obs})\,,\nn
\end{eqnarray}
where we have explicitly kept the dependence on the observer spacetime coordinates $(\eta_{\obs}, \bx_{\obs})$. The perturbation $\mathcal{E}(\eta_{\obs}, \bx_{\obs},  \bee, \nu_{\obs})$  is explicitly given by
%\begin{widetext}
% \begin{eqnarray}\label{sss}
%&&\mathcal{E}(\eta_{\obs}, \bx_{\obs},  \bee, \nu_{\obs})=\frac{1}{4\pi}\int \dd\eta \,a^4 \int d\theta_{\Gal}\, \bar{n}_{\Gal}\,\mathcal{L}_{\Gal}(\bar{\nu}_{\Gal}, \theta_{\Gal})\times \\
%&\times&\left\{b\delta_{\rm CDM}+4\Psi+\textcolor{black}{4}\Pi-2\bee\cdot \nabla v-6\int_{\eta_{\obs}}^{\eta}\dd\eta' \dot{\Psi}+\frac{1}{\mathcal{L}_{\Gal}(\bar{\nu}_{\Gal}, \theta_{\Gal})}\frac{\partial \mathcal{L}_{\Gal}}{\partial \nu_{\Gal}}(\nu_{\Gal}, \theta_{\Gal})\Big|_{\bar{\nu}_{\Gal}}\frac{\nu_{\obs}}{a}\left[\bee\cdot \nabla v-\Psi-\Pi+2\int_{\eta_{\obs}}^{\eta} d\eta' \dot{\Psi}\right]\right\}\,,\nn
%\end{eqnarray}
%\end{widetext}
\begin{widetext}
 \begin{eqnarray}\label{sss}
&&\mathcal{E}(\eta_{\obs}, \bx_{\obs},  \bee, \nu_{\obs})=\frac{1}{4\pi}\int \dd\eta \,a^4 \int d\theta_{\Gal}\, \bar{n}_{\Gal}\,\mathcal{L}_{\Gal}(\bar{\nu}_{\Gal}, \theta_{\Gal})\times \\
&\times&\left\{b\delta_{\rm CDM}+4\Psi+\textcolor{black}{4}\Pi-2\bee\cdot \nabla v-6\int_{\eta_{\obs}}^{\eta}\dd\eta' \dot{\Psi}+\frac{1}{\mathcal{L}_{\Gal}}\frac{\partial \mathcal{L}_{\Gal}}{\partial \nu_{\Gal}}\Big|_{{\bar{\nu}}_{\Gal}}\frac{\nu_{\obs}}{a}\left[\bee\cdot \nabla v-\Psi-\Pi+2\int_{\eta_{\obs}}^{\eta} d\eta' \dot{\Psi}\right]\right\}\,,\nn
\end{eqnarray}
\end{widetext}
where we have set to zero all quantities at the observer, since they do not depend on $\bee$ and thus do not contribute to the anisotropy and are a correction to the monopole of the GW background.  This expression can be compared to Eq. (9) of Ref. \cite{Contaldi:2016koz} and gives the expression of the emissivity in terms of the astrophysical details of the galaxies and thus its frequency dependence.

In Eqs. (\ref{ss}) and (\ref{sss}) we distinguish two different types of contributions: (1) a contribution from local physics, proportional to the galaxy effective luminosity $\mathcal{L}_{\Gal}$ and (2) a cosmological contribution which depends on metric perturbations, matter density contrast and velocities. In the cosmological part, we can distinguish different types of terms: a contribution propositional to the CDM overdensity, coming from the perturbative expansion of the galaxy density $n_{\Gal}$ in the covariant expression (\ref{mastermaster}); a Doppler contribution due to the peculiar motion of galaxies with respect to the observer rest frame; local contributions and a contribution integrated along the line of sight. We expect the local contribution proportional to the CDM overdensity to be the dominant term, together with the Doppler one. A quantitative analysis of this expression and an estimate of its various contributions will be presented in \cite{Cusinprep}.  

We conclude this section observing that the cosmological factor in eq. (\ref{sss}) has a structure very similar to the one of the Sachs-Wolfe formula for CMB temperature anisotropies. Nevertheless, this result is absolutely not trivial and could not be guessed from its electromagnetic counterpart. Indeed, it is true that in both cases (CMB and AGWB) a line of sight approach is exploited together with the eikonal approximation. However, for the electromagnetic case, photons are coming to the observer from a constant time hypersurface, while in the case under study, sources emitting GW have a non trivial redshift distribution. As a consequence, in this gravitational situation, when calculating (\ref{ss}), we had to take into account an effect of volume distortion, absent in the CMB case~\footnote{In deriving Eqs. (\ref{mastermaster}) and (\ref{ss}) the reciprocity relation (\ref{reciprocity}) has been exploited and it allowed to get considerable computational simplifications.}.

 %%%%%%%%%%%%%%%%%%%%%%%%%%%%%%%%%%%
\section{Angular power spectrum}\label{angular}\label{angular}
%%%%%%%%%%%%%%%%%%%%%%%%%%%%%%%%%%%

The function $\mathcal{E}(\eta_{\obs}, \bx_{\obs}, \nu_{\obs}, \bee)$ defined in Eq. (\ref{sss}) is a stochastic variable which depends on the observer's position and direction of observation. The cosmological variables are correlated stochastic variables whose spectra are related to a scenario of structure formation~\cite{PeterUzan2005}. It follows that $\mathcal{E}$ can be characterized by its angular correlation function
\be
C(\theta)\equiv \langle   \mathcal{E}(\eta_{\obs}, \bx_{\obs}, \nu_{\obs}, \bee_1)  \mathcal{E}(\eta_{\obs}, \bx_{\obs}, \nu_{\obs}, \bee_2)\rangle\,.
\ee
Statistical isotropy implies that the correlation function depends only on the relative angle $\theta$; $\bee_1\cdot \bee_2 \equiv\cos\theta$. This correlation function can be expanded in Legendre polynomials 
\be\label{above}
 C(\theta)\equiv \sum_{\ell}\frac{2\ell+1}{2\pi} C_{\ell} P_{\ell}(\bee_1\cdot \bee_2)\,.
\ee
A given multipole $\ell$ corresponds to the typical angular scale $\pi/\theta$ and $C_{\ell}$ is an estimate of the variance of GW energy density fluctuations on that scale. To get its expression, we first decompose $\mathcal{E}$  in spherical harmonics as
\be\label{apre}
\mathcal{E}(\eta_{\obs}, \bx_{\obs}, \nu_{\obs}, \bee)=\sum_{\ell m} a_{\ell m} (\eta_{\obs}, \bx_{\obs}, \nu_{\obs}) Y_{\ell m}(\bee)\,,
\ee
where the coefficients of the development are given by
\be\label{a}
a_{\ell m}(\eta_{\obs}, \bx_{\obs}, \nu_{\obs})\equiv \int \dd^2\bee\, \mathcal{E}(\eta_{\obs}, \bx_{\obs}, \bee, \nu_{\obs}) Y_{\ell m}^*(\bee)\,.
\ee
Making use of the properties of Legendre polynomials, it is easy to show that 
\be
(2\ell+1) C_{\ell}\equiv \sum_m \langle a_{\ell m} a^*_{\ell m}\rangle\,\label{CR},
\ee
where the brackets stand for an ensemble average over the stochastic variables. Statistical homogeneity implies that this shall not depend on $\bx_{\obs}$. From now on we will thus omit the dependence of $a_{\ell m}$ on quantities at the observer. 

In order to calculate $C_{\ell}$, it is useful to first expand $\mathcal{E}$ in Fourier modes as
\be\label{fR}
\mathcal{E}(\eta_{\obs}, \bx_{\obs}, \nu_{\obs}, \bee)=\int \frac{\dd^3\bk}{(2\pi)^3}  \mathcal{\hat{E}}_\bk(\eta_{\obs}, \nu_{\obs}, \bee)\,,\\
\ee
where we have reabsorbed a phase $\exp(i \bk \cdot \bx_{\obs})$ into the definition of the field $\mathcal{\hat{E}}$.  All variables are decomposed in a similar way and with the same convention for the phase $\exp(i \bk \cdot \bx_{\obs})$. Since $\mathcal{E}$ is an integral over the background geodesic $\bx(\eta)=\bx_{\obs}+\bee\left(\eta_{\obs}-\eta\right)$, we get
\begin{widetext}
\begin{eqnarray}\label{Rk}
&&\mathcal{\hat{E}}_\bk(\eta_{\obs}, \bee, \nu_{\obs})=
 \frac{1}{4\pi}\int \dd\eta \,a^4 \int \dd\theta_{\Gal}\, \mathcal{L}_{\Gal}(\bar{\nu}_{\Gal}, \theta_{\Gal}) \,\bar{n}_{\Gal}(\eta, \theta_{\Gal})\times \nn\\
 &&\times \left\{\hbox{e}^{i k \mu \Delta\eta}
 \left[4\hat{\Psi}_\bk(\eta)+4\hat{\Pi}_\bk(\eta)+b\hat{\delta}_\bk(\eta)-2i k\mu\, \hat{v}_\bk(\eta)+\frac{1}{\mathcal{L}_{\Gal}}\frac{\partial \mathcal{L}_{\Gal}}{\partial \nu_{\Gal}}\Big|_{\bar{\nu}_{\Gal}}\frac{\nu_{\obs}}{a}\left( -\hat{\Psi}_\bk(\eta)-\hat{\Pi}_\bk(\eta)+i k\mu\, \hat{v}_\bk(\eta)  \right) \right]+\right.\nn\\
 &&\left.\quad +2\left(\frac{1}{\mathcal{L}_{\Gal}}\frac{\partial \mathcal{L}_{\Gal}}{\partial \nu_{\Gal}}\Big|_{\bar{\nu}_{\Gal}}\frac{\nu_{\obs}}{a}-3\right) \int_{\eta_{\obs}}^{\eta} \dd\eta' \hbox{e}^{ik\mu\Delta \eta'} \dot{\hat{\Psi}}_\bk(\eta')\right\}\,,
\end{eqnarray}
\end{widetext}
where $\bee\cdot \bk=k\mu$. Each stochastic variable $\hat X_\bk(\eta)$ in Eq.~(\ref{Rk}) can be decomposed as the product of a transfer function $X_k(\eta)$ and a unique random variable $\hat a_\bk$ as
\be
\hat X_\bk(\eta)=\hat X_k(\eta) \hat a_\bk\,.
\ee
$\hat{X}_k(\eta)$ depends only on the modulus of $\bk$  and $\hat a_\bk$ satisfies 
\be
\langle \hat a_\bk \hat a_{\bk'}^*\rangle =(2\pi)^3 \delta^{(3)} (\bk-\bk')\,.
\ee
It follows that the mode function of ${\cal\hat E}$ can be decomposed as
\be\label{1111}
\mathcal{\hat{E}}_\bk(\eta_{\obs}, \nu_{\obs}, \bee)\equiv\,\mathcal{\hat{E}}_k(\eta_{\obs}, \nu_{\obs}, \mu)\hat a_\bk\,.\\
\ee
We now expand the exponentials in spherical harmonics and spherical Bessel functions using the standard relation
\be\label{magic}
\hbox{e}^{i \bp\cdot \bx}=4\pi\sum_{\ell m} i^{\ell} j_{\ell} (px)Y_{\ell m}^*(\hat{\bp}) Y_{\ell m}(\hat{\bx})\,.
\ee
With the definition 
\begin{align}\label{Rkkk}
\mathcal{\hat{E}}_k(\eta_{\obs}, \nu_{\obs}, \mu)&=4\pi \sum_{\ell m} i^{\ell}\,\hat{\mathcal{E}}_{\ell}(k,\nu_{\obs}) Y_{\ell m}^*(\hat{\bk}) Y_{\ell m}(\bee)\,,
\end{align}
%where again, $\bee\cdot \bk=k\mu$, 
we finally get
  \begin{widetext}
\begin{eqnarray}\label{Rkk}
&&\hat{\mathcal{E}}_{\ell}(k,\nu_{\obs})=\frac{1}{4\pi}\int \dd\eta \,a^4 \int \dd\theta_{\Gal}\,\mathcal{L}_{\Gal}(\bar{\nu}_{\Gal}, \theta_{\Gal}) \,\bar{n}_{\Gal}(\eta, \theta_{\Gal})\times\nn \\
&&\times  \left\{\left[4\hat{\Psi}_k(\eta)+4\hat{\Pi}_k(\eta)+b\hat{\delta}_k(\eta)\right] j_{\ell}(k\Delta\eta)-2k \hat{v}_k(\eta)j'_{\ell}(k\Delta\eta)-6\int_{\eta_{\obs}}^{\eta} \dd\eta'\dot{\hat{\Psi}}_k(\eta') j_{\ell}(k\Delta\eta')\right.+\nn\\
&&+\left.\frac{1}{\mathcal{L}_{\Gal}}\frac{\partial \mathcal{L}_{\Gal}}{\partial \nu_{\Gal}}\Big|_{\bar{\nu}_{\Gal}}\frac{\nu_{\obs}}{a}  \left[-\hat{\Psi}_k(\eta)j_{\ell}(k\Delta\eta)-\hat{\Pi}_k(\eta)j_{\ell}(k\Delta\eta)+k \hat{v}_k(\eta)j'_{\ell}(k\Delta\eta)+2\int_{\eta_{\obs}}^{\eta} \dd\eta'\dot{\hat{\Psi}}_k(\eta') j_{\ell}(k\Delta\eta')\right]\right\}\,,
\end{eqnarray}
\end{widetext}
where the prime acting on the spherical Bessel denotes a derivative with respect to its argument. 

We can now go back to the expression (\ref{CR}) defining the angular power spectrum $C_{\ell}$.  Plugging in Eq. (\ref{a})  defining the multipoles $a_{\ell m}$, the Fourier transform (\ref{fR}) of $\mathcal{E}$ and using Eqs.~(\ref{1111}) -(\ref{Rkk}), we get
\be\label{Cell}
C_{\ell}(\nu_{\obs})=\frac{2}{\pi}\int \dd k\,k^2 |\mathcal{E}_{\ell}(k,\nu_{\obs})|^2\,.
\ee
This expression is similar to the ones of CMB temperature and polarisation angular power spectrum. Note however that we have one angular spectrum per frequency band. 

From the result (\ref{Cell}) we can compute the variance of $\Omega_{\rm GW}$ due to the distribution of the large scale structures. For each frequency band, the variance is defined by 
\begin{eqnarray}
\sigma^2_{\text{GW}}(\nu_{\obs})&\equiv&\langle \delta\rho_{\text{GW}}(\nu_{\obs}, \bee_1)\delta\rho_{\text{GW}}(\nu_{\obs}, \bee_2)\rangle\vline_{\,\bee_1\cdot\bee_2=1}\,,\nn\\
%&=&C(\vartheta=0)\,,
\end{eqnarray}
so that using Eq. (\ref{above}), we get 
\be
\sigma^2_{\text{GW}}(\nu_{\obs})\equiv \sum_{\ell}\frac{(2\ell+1)}{2\pi}C_{\ell}(\nu_{\obs})\,.
\ee
The computation of $\sigma^2_{\text{GW}}(\nu_{\obs})$ requires to solve the evolution of the perturbations and of the galaxy number counts (via a merger-tree analysis). The evolution of the cosmological perturbations can be split in terms of a transfer function and an initial power spectrum. When the dominant contribution to (\ref{Cell}) is proportional to $\delta_{\text{CDM}}$ and the initial power spectrum is scale invariant, it takes the form
\begin{widetext}
\be
\sigma^2_{\text{GW}}(\nu_{\obs})=\sum_{\ell}\frac{2\ell+1}{16\pi^4}\int \frac{dk}{k}\left[\int d\eta \,a^4(\eta) \int d\theta_{\Gal}\mathcal{L}(\bar{\nu}_{\Gal}, \theta_{\Gal})\bar{n}_{\Gal}(\eta, \theta_{\Gal})b(\eta) T_{\delta}(k, \eta)j_{\ell}(k\Delta\eta)\right]^2\,, 
\ee
\end{widetext}
where $T_{\delta}(k, \eta)$ is the transfer function of matter density contrast. Approximate analytic expressions for $\sigma^2_{\text{GW}}(\nu_{\obs})$ and its numerical computation will be presented in our companion article \cite{Cusinprep}.

\section{Correlation with lensing}\label{lensing}

The GW energy density~(\ref{master}) depends on the local properties of galaxies as well as on cosmological perturbations. This means that it will be correlated with other probes and in particular weak lensing of background galaxies and galaxy number counts.

Weak lensing describes the deformation of the shape of background galaxies by the gravitational potential of the large scale structure. Its description is now standard, see e.g. Refs.~\cite{PeterUzan2005,1992grle.bookS,Bartelmann:1999yn,Fleury:2015rwa}. The observed distortion in a given direction is characterized by the Jacobi matrix ${\bm{\mathcal{D}}}$ and the associated amplification matrix ${\bm{\mathcal{A}}}$, that relates the observed and intrinsic shapes.

In a perturbed FLRW spacetime, the amplification matrix splits into a background  and a perturbed contribution as ${\bm{\mathcal{A}}}={\bm{\mathcal{A}}}^{(0)}+{\bm{\mathcal{A}}}^{(1)}$. The background part reduces to ${\bm{\mathcal{A}}}^{(0)}={\bf{1}}$ and the first order contribution is
\be
\left({\bm{\mathcal{A}}}^{(1)}\right)_{ab}(\eta, \bee)=-\partial_a\partial_b \varphi\,,
\ee
with the partial derivatives defined on the 2-sphere and 
\be\label{effc}
\varphi(\chi, \bee)\equiv\frac{2}{c^2}\int_0^{\chi}d\chi'\, \frac{(\chi-\chi')}{\chi\chi'} \Psi(\bee, \chi') d\chi'\,,
\ee
where the integral runs along the unperturbed geodesic $\bx=\bx_{\obs}+\bee(\eta_{\obs}-\eta)$.  The convergence $\kappa$ is defined as the half of the trace of the amplification matrix while its trace-free part defines the cosmic shear $(\gamma_1,\gamma_2)$. The total convergence is obtained by integrating this contribution weighted by the number of sources along the line of sight, as
\be
\kappa^{(1)}(\bee)=\frac{1}{c^2}\int_0^{\chi_H}d\chi\, g(\chi)\, \Delta_2 \Psi(\bee, \chi)\,,
\ee
with 
\be
g(\chi)\equiv\frac{1}{\chi} \int_{\chi}^{\chi_H}d\chi' p_{\chi}(\chi') \frac{(\chi'-\chi)}{\chi'}\,.
\ee
and $\chi_H$ is the value of $\chi$ associated to the size of the observable universe.
%\textcolor{red}{stop}
%
%
%In a perturbed FLRW spacetime, the amplification matrix splits into a background  and a perturbed contribution as ${\bm{\mathcal{A}}}={\bm{\mathcal{A}}}^{(0)}+{\bm{\mathcal{A}}}^{(1)}$. The background part reduces to ${\bm{\mathcal{A}}}^{(0)}={\bf{1}}$ and the first order contribution is
%\be
%\left({\bm{\mathcal{A}}}^{(1)}\right)_{ab}(\eta, \bee)=-2\int_{\eta_{\obs}}^{\eta} \dd\eta' \frac{\eta' (\eta-\eta')}{\eta} \partial_a\partial_b \Psi\,,
%\ee
%where the integral runs along the unperturbed geodesic $\bx=\bx_{\obs}+\bee(\eta_{\obs}-\eta)$.  The trace of the amplification matrix is the convergence $\kappa$ while its tracefree part defines the cosmic shear $(\gamma_1,\gamma_2)$. The convergence is given by
%\be
%\kappa^{(1)}(\eta, \bee)=-2\int_{\eta_{\obs}}^{\eta} d\eta' \frac{\eta' (\eta-\eta')}{\eta} \Delta \Psi\,
%\ee
%for a source located at $\eta$. The total convergence is obtained by integrating this contribution weighted by the number of sources along the line of sight,
%\be\label{effc}
%\kappa^{(1)}(\bee)=-2\int \dd\eta\, p_{\eta}(\eta)\int_{\eta_{\obs}}^{\eta} \dd\eta' \frac{\eta' (\eta-\eta')}{\eta} \Delta \Psi\,,
%\ee
%where $p_{\eta}$ is a function that takes into account the redshift distribution of sources, $\dd\eta\, p_{\eta}(\eta)=\dd z\, p_z(z)$. Eq.~(\ref{effc}) can be rewritten as
%\be
%\kappa^{(1)}(\bee)=-2 \int \dd\eta\,g(\eta)\eta \, \Delta\Psi \,,
%\ee
%with
%\be
%g(\eta)\equiv \int_{\eta}^{\eta_H}\dd\eta'\,p_{\eta}(\eta')\frac{\eta'-\eta}{\eta'}\,,
%\ee
%and $\eta_H$ is the value of $\eta$ associated to the size of the observable universe. 
In this analysis, we just consider the total convergence, but our result can be trivially extended to redshift bins if a tomographic information exists.\\

The convergence~(\ref{effc}) depends on the lensing potential $2\Psi$ and on the direction of observation. Since the GW energy density also depends on the perturbation variables, both quantities are correlated. We introduce the cross-correlation 
\begin{align}\label{B}
B(\theta,\nu_{\obs})&\equiv \langle \mathcal{E}(\eta_{\obs},\nu_{\obs}, \bx_{\obs}, \bee_1, \nu_{\obs}) \kappa(\eta_{\obs}, \bx_{\obs}, \bee_2)\rangle\,,
\end{align}
where $\bee_1\cdot \bee_2=\cos\theta$. This function can be expanded in Legendre polynomials as
\be
B(\theta,\nu_{\obs})=\sum_{\ell} \frac{2\ell+1}{2\pi} B_{\ell}(\nu_{\obs})\,P_{\ell}(\bee_1\cdot \bee_2)\,,
\ee
with $\bee_1\cdot \bee_2=\cos\theta$. Following the same steps as in section \ref{angular}, we expand $\kappa$ in multipoles as
\be\label{dpre}
\kappa(\eta_{\obs}, \bx_{\obs}, \bee)=\sum_{\ell m}\kappa_{\ell m}(\eta_{\obs}, \bx_{\obs}) Y_{\ell m}(\bee)\,,
\ee
with 
\be\label{d}
\kappa_{\ell m}(\eta_{\obs}, \bx_{\obs})=\int \dd^2\bee\, \kappa(\eta_{\obs}, \bx_{\obs}, \bee) Y_{\ell m}^*(\bee)\,.
\ee
Inserting these decompositions together with Eqs. (\ref{apre}-\ref{a}) and (\ref{dpre}-\ref{d}) for the multipolar decomposition of  ${\cal E}$ in Eq. (\ref{B}), we get 
\be\label{Bl}
(2\ell+1)B_{\ell}(\nu_{\obs})=\sum_m \langle a_{\ell m}(\nu_{\obs}) \kappa_{\ell m}^*\rangle\,.
\ee
The calculation of this quantity follows the same line as in section~\ref{angular}: one needs to expand $\mathcal{E}(\eta_{\obs}, \bx_{\obs}, \nu_{\obs}, \bee)$ and $\kappa(\eta_{\obs}, \bx_{\obs}, \bee)$ in Fourier modes, hence defining
\be\label{fK}
\hat{\kappa}_\bk(\eta_{\obs}, \bee)=2k^2 \int \dd\eta \,g(\eta)\,\eta\, \hat{\Psi}_\bk(\eta') \hbox{e}^{ik\mu\Delta\eta'}\,.
\ee
Decomposing the Bardeen potential in terms of a transfer function $\Psi_k(\eta)$ and the unit random variable $\hat a_\bk$, one gets
\be\label{KK}
\hat{\kappa}_\bk(\eta_{\obs}, \bee)= \hat{\kappa}_k(\eta_{\obs}, k, \mu)\hat a_\bk\,.
\ee
After decomposing the exponential in spherical harmonics, we get 
\be\label{KKK}
 \hat{\kappa}_k(\eta_{\obs}, \mu)\equiv 4\pi\sum_{\ell m} i^{\ell} \hat{\kappa}_{\ell}(k) Y_{\ell m}^*(\hat{\bk}) Y_{\ell m}(\bee) \,,
\ee
with
\be
\hat{\kappa}_{\ell}(k)=-\ell(\ell+1)\frac{1}{c^2}\int_0^{\chi_H}d\chi\, g(\chi)\, \hat{\Psi}_k(\chi) j_{\ell}(k\chi)\,,
\ee
or similarly
\be
\hat{\kappa}_{\ell}(k)=\ell(\ell+1)\frac{1}{c^2}\int_{\eta_{\obs}}^{\eta_H}d\eta\, g(\eta)\,  \hat{\Psi}_k(\eta) j_{\ell}(k\Delta\eta)\,.
\ee
%
%\be
%\hat{\kappa}_{\ell}(k)=2k^2 \int \dd\eta\, g(\eta)\, \eta\, j_{\ell}(k\Delta\eta) \hat{\Psi}(\eta, k)\,.
%\ee
To conclude, we just need to substitute the expressions of the Fourier decompositions of $\mathcal{E}$ and $\kappa$, i.e. Eqs. (\ref{Rkkk}-\ref{Rkk}) and (\ref{KK}-\ref{KKK}), in the definitions (\ref{a}) and (\ref{d})  of the multipoles $a_{\ell m}$ and $\kappa_{\ell m}$, to get
\be\label{Bell}
B_{\ell}(\nu_{\obs})=\frac{2}{\pi}\int \dd k\,k^2 \mathcal{E}^*_{\ell}(k,\nu_{\obs})\,\kappa_{\ell}(k)\,.
\ee

\section{Correlation with galaxy number counts}\label{counts}

Galaxy surveys provide catalogs with information about galaxy redshift $z_{\Gal}$, angular coordinates, fluxes and shapes. Redshift space coordinates need to be converted into physical coordinates, with which theoretical predictions are drawn.

The galaxy number counts ~\cite{Bonvin:2011bg,Challinor:2011bk}, defined as the perturbation of the number of galaxies as a function of direction and redshift\footnote{Being an observable quantity and therefore gauge invariant, we have the freedom to derive it in an arbitrary gauge.}, is also correlated with the GW background. We consider the number of galaxies in a direction $\bee$ at redshift $z_{\Gal}$, that we denote as 
\be
N_{\Gal}(\bee, z_{\Gal})\dd^2\Omega \dd z_{\Gal}\,.
\ee
The redshift distribution $\langle N_{\Gal}\rangle (z_{\Gal}) \dd z_{\Gal}$ is deduced from this quantity taking an average over the sky. The galaxy density perturbation at a fixed redshift $z_{\Gal}$ observed in a direction $\bee$ is given by
\begin{eqnarray}
\delta_z(\bee, z_{\Gal}) &\equiv& \frac{n_{\Gal}(\bee, z_{\Gal})-\langle n_{\Gal}\rangle (z_{\Gal})}{\langle n_{\Gal}\rangle (z_{\Gal})}\\
&\equiv& \frac{N_{\Gal}(\bee, z_{\Gal})-\langle N_{\Gal}\rangle (z_{\Gal})}{\langle N_{\Gal}\rangle (z_{\Gal})}-\frac{\delta \nu(\bee, z_{\Gal})}{\nu(z_{\Gal})}\,,\nn
\end{eqnarray}
where we have introduced the galaxy number density 
\be
n_{\Gal}(\bee, z_{\Gal})\equiv\frac{N_{\Gal}(\bee, z_{\Gal})}{\nu(\bee, z_{\Gal})}\,,
\ee
and the physical survey volume density,  $\nu$, per redshift bin and solid angle.  $\nu$ is also a perturbed quantity since the solid angle of observation as well as the redshift bin are distorted between the source and the observer. Expressing the physical volume as $V(\bee, z_{\Gal})\equiv \nu(\bee, z_{\Gal}) \dd z_{\Gal} \dd^2\Omega$, it follows that
\be
\nu(\bee, z_{\Gal})\equiv \nu(z_{\Gal})+\delta \nu(\bee, z_{\Gal})\,.
\ee
We can therefore define the galaxy number counts as
\begin{align}
\Delta (\bee, z_{\Gal})&\equiv \frac{N_{\Gal}(\bee, z_{\Gal})-\langle N_{\Gal}\rangle (z_{\Gal})}{\langle N_{\Gal}\rangle (z_{\Gal})}\\
&\equiv \delta_z(\bee, z_{\Gal})+\frac{\delta\nu(\bee, z_{\Gal})}{\nu(z_{\Gal})}\,,
\end{align}
in terms of redshift perturbation $\delta_z$ and volume density perturbation $\delta\nu/\nu$. All these quantities are gauge invariant. An explicit calculation of $\Delta$ in terms of cosmological quantities can be  found in Ref.~\cite{Bonvin:2011bg}. 

In the following, we consider the most dominant term of the galaxy number counts (beside the intrinsic density contrast), following the approximation first physically motivated by Kaiser \cite{Kaiser:1987qv}. Since then it has been considered to estimate the effect of peculiar velocities on the observed redshift. In the following, we refer to FLRW comoving coordinate frame as real space and to the observed coordinates as redshift space, indicated with subscript ${\rm obs}$. Galaxy number conservation must bring the same result in real space and redshift space
\be\label{cons}
n_{\Gal, \text{obs}}(\bx_{\text{obs}})\dd^3\bx_{\text{obs}}=n_{\Gal}(\bx)\dd^3\bx\,.
\ee
We now define 
\begin{align}
n_{\Gal, \text{obs}}(\bx_{\text{obs}})&\equiv \bar{n}_{\Gal}\left[1+\delta_{\text{obs}}(\bx_{\text{obs}})\right]\,,\\
n_{\Gal}(\bx)&\equiv \bar{n}_{\Gal}\left[1+\delta(\bx)\right]\,,
\end{align}
where we have used that in the absence of perturbations, we have no space distortion. From Eq. (\ref{cons}), it follows
\be\label{isolate}
\left[1+\delta_{\Gal, \text{obs}}(\bx_{\text{obs}})\right]\dd^3\bx_{\text{obs}}=\left[1+\delta_{\Gal}(\bx)\right]\dd^3\bx\,.
\ee
To isolate $\delta_{\Gal, \text{obs}}(\bx_{\text{obs}})$ we need to calculate the Jacobian of the transformation from redshift space coordinates to real space. Since only the radial coordinate is affected by redshift perturbations, we have
\be
\chi_{\text{obs}}\equiv \chi(z_{\Gal})\simeq \chi(\bar{z})+\frac{\partial \chi}{\partial \bar{z}}\delta z\,,
\ee
where  we have used $z_{\Gal}=\bar{z}+\delta z$ and $\partial \chi/\partial \bar{z}=\mathcal{H}^{-1}(1+\bar{z})^{-1}$. Keeping only the Doppler contribution in Eq. (\ref{expz}), we find 
\be
\chi_{\text{obs}}\simeq \chi+\frac{\bee\cdot \bv}{\mathcal{H}}\,.
\ee
The Jacobian, 
\begin{align}
J &\equiv \left| \frac{\dd^3\bx}{\dd^3\bx_{\text{obs}}}\right|=\frac{\dd\chi}{\dd\chi_{\text{obs}}}\frac{\chi^2}{\chi^2_{\text{obs}}}\,,
\end{align}
reduces at linear order to
\begin{align}
J &\simeq 1-\frac{\partial}{\partial \chi}\left[ \frac{\bee\cdot \bv}{\mathcal{H}}\right]-2\frac{\bee\cdot \bv}{\mathcal{H}\chi}\,,
\end{align}
where we have also kept the dominant term in the limit $|\bv|\rightarrow 0$. Going back to Eq.~(\ref{isolate}), it follows
\be
\delta_{\Gal, \text{obs}}(\bx)=\delta_{\Gal}(\bx)+\frac{\partial}{\partial \eta}\left[ \frac{\bee\cdot \bv}{\mathcal{H}}\right]-2\frac{\bee\cdot \bv}{\mathcal{H}\chi}\,.
\ee
The last term is usually neglected, since the derivative of perturbations are expected to dominate \footnote{Only on very large scales comparable to the Hubble radius $H^{-1}$, the contribution of the last term is non-negligible and should be carefully considered when interested in wide-angle surveys \cite{Matsubara:2004fr, Montanari:2015rga, Matsubara:1999du,Reimberg:2015jma}.} together with the dominant intrinsic clustering term \cite{Green:2014aga}. Identifying the observed number counts with the redshift-space overdensity, we obtain the so-called  Kaiser approximation for redshift-space distortions 
\be
\Delta(\bx)\simeq b\, \delta_{\rm CDM}(\bx)+\frac{\partial}{\partial \eta}\left[ \frac{\bee\cdot \bv}{\mathcal{H}}\right]\,, 
\ee
where $b$ is the bias (that we consider to be a function of time) and $v$ the cold dark matter velocity flow.  We can integrate along the line of sight to get the effective number count, function of the direction of observation,
\be
\Delta(\eta_{\obs}, \bx_{\obs}, \bee)\equiv \int d\eta \Delta(\eta_{\obs}, \bx_{\obs}, \bee, \eta)\,,
\ee
where we have explicitly indicated the dependence on the observer spacetime coordinates.  Explicitly, in the Kaiser approximation
\be\label{Kaiser}
\Delta(\eta_{\obs}, \bx_{\obs}, \bee)\simeq \int \dd\eta\left[b\, \delta_{\rm CDM}(\bx)+\frac{\partial}{\partial \eta}\left( \frac{\bee\cdot \bv}{\mathcal{H}}\right)\right]\,.
\ee

Let us now turn to the correlation between $\Delta$ and the GW energy density defined in Eq.~(\ref{ss}). It is given by
\begin{align}\label{D}
D(\theta,\nu_{\obs})&\equiv \langle \mathcal{E}(\eta_{\obs}, \bx_{\obs}, \bee_1, \nu_{\obs}) \Delta(\eta_{\obs}, \bx_{\obs}, \bee_2)\rangle,
\end{align}
with $\bee_1\cdot \bee_2=\cos\theta$. Again, it can be expanded in Legendre polynomials as
\be
D(\theta,\nu_{\obs})=\sum_{\ell} \frac{2\ell+1}{2\pi} D_{\ell}(\nu_{\obs})\,P_{\ell}(\bee_1\cdot \bee_2)\,.
\ee
Following the same procedure as in section \ref{angular}, we expand $\Delta$ in multipoles as
\be\label{bpre}
\Delta(\eta_{\obs}, \bx_{\obs}, \bee)=\sum_{\ell m}d_{\ell m}(\eta_{\obs}, \bx_{\obs}) Y_{\ell m}(\bee)\,,
\ee
with 
\be\label{b}
d_{\ell m}(\eta_{\obs}, \bx_{\obs})=\int \dd^2\bee\, \Delta(\eta_{\obs}, \bx_{\obs}, \bee) Y_{\ell m}^*(\bee)\,,
\ee
so that
\be\label{Dl}
(2\ell+1)D_{\ell}(\nu_{\obs})=\sum_m \langle a_{\ell m}(\nu_{\obs}) d_{\ell m}^*\rangle\,.
\ee
In order to calculate $D_\ell$, we expand $\Delta(\eta_{\obs}, \bx_{\obs}, \bee)$ in Fourier space as
\be
\Delta(\eta_{\obs}, \bx_{\obs}, \bee)=\int \frac{\dd^3\bk}{(2\pi)^3} \hat{\Delta}_\bk(\eta_{\obs}, \bee)\,, 
\ee
the phase $\exp(i\bk\cdot \bx_{\obs})$ being absorbed in the definition of the Fourier coefficient. We find 
\begin{align}\label{fD}
& \hat{\Delta}_\bk(\eta_{\obs},  \bee)=\\
&=\int \dd\eta \left[b\hat{\delta}_{{\rm CDM}\bk}(\eta)+ik\mu \partial_{\eta}\left(\frac{\hat{v}_\bk(\eta)}{\mathcal{H}}\right)\right]e^{ik\mu \Delta\eta}\,,\nn
\end{align}
with $\bee\cdot \bk=k\mu$. Decomposing  $\hat{\bv}_\bk(\eta)$ and $\hat{\delta}_{{\rm CDM},\bk}(\eta)$ as the product of a random variable $\hat a_\bk$ and transfer function $\hat{\bv}_k(\eta)$ and $\hat\delta_{{\rm CDM},\bk}$, we get
\be\label{DD}
\hat{\Delta}_\bk(\eta_{\obs}, \bee)=\hat a(\bk) \hat{\Delta}_k(\eta_{\obs}, \mu)\,.
\ee
Expanding the exponential functions in spherical harmonics by using Eq.~(\ref{magic}), we can define
\be\label{DDD}
 \hat{\Delta}_k(\eta_{\obs}, \mu)\equiv 4\pi\sum_{\ell m} i^{\ell} \hat{\Delta}_{\ell}(k) Y_{\ell m}^*(\hat{\bk}) Y_{\ell m}(\bee) \,,
\ee
with
\begin{align}
&\hat{\Delta}_{\ell}(k)=\\
&=\int \dd\eta\, \left[b\hat{\delta}_{{\rm CDM},k}(\eta) j_{\ell}(k\Delta\eta)+k \partial_{\eta}\left(\frac{\hat{v}_k}{\mathcal{H}}\right)j'_{\ell}(k\Delta\eta)\right]\,.\nn
\end{align}
The cross-correlation is then obtained exactly in the same way as in the previous section, so that
\be\label{Dell}
D_{\ell}(\nu_{\obs})=\frac{2}{\pi}\int \dd k\,k^2 \mathcal{E}^*_{\ell}(k,\nu_{\obs})\,\Delta_{\ell}(k)\,.
\ee

%%%%%%%%%%%%%%%%%%%%%%%%%%%%%%%%%%%%%%%%%
\section{Conclusions}\label{conclusions}
%%%%%%%%%%%%%%%%%%%%%%%%%%%%%%%%%%%%%%%%%

We have developed an analytic approach for the computation of the anisotropies of the background of gravitational radiation from unresolved galactic and extragalactic sources. We have distinguished three different scales in the problem: a cosmological, a galactic, and an astrophysical scale. We have explained how to relate quantities measured by the observer (i.e. the directional flux of gravitational radiation per unit of frequency) to effective and local quantities characterizing the GW emission. At each scale, different mathematical tools need to be exploited: from cosmological perturbation theory at the cosmological scale, to statistical mechanics and astrophysical calculation to characterize the local process of GW emission. This coarse graining approach enables us to obtain a parametrization for the observed flux of GW in a given direction, as a function of local quantities characterizing the process of GW emission and already presented in the literature, for various source types. 

Our final results for the anisotropies of the AGWB energy density, contains a cosmological component, which depends on the distribution of structures at large scales, and a local one, related to the local process of GW emission and to the intrinsic properties of the sources, see. Eq. (\ref{mastermaster}) for the general expression for  AGWB anisotropies and eq. (\ref{ss}) for the expression specialized to a FLRW universe with scalar perturbations. We have derived the expression of the AGWB angular power spectrum, see Eq. (\ref{Cell}) as well as its correlation with lensing and galaxy number counts, Eqs. (\ref{Bell}) and (\ref{Dell}), respectively. 

In the next few decades the LISA and ET experiments will provide new data on an unexplored frequency band, where many galactic and extragalactic astrophysical processes are expected to contribute. In principle, observations will allow us to reconstruct an angular map of the AGWB energy density, similar to the CMB map for electromagnetic radiation. On the other hand, modeling the astrophysical background as precisely as possible to extract information on its strength, frequency range and statistical properties, is crucial to distinguish it from the cosmological signal, foreground  or instrumental noise. 

Even though this observational landscape may seem far away, our analysis (and its companion in \cite{Cusinprep}) will enable us to quantify the relative influences of the astrophysical and cosmological sources. It will either open a window on astrophysics (e.g. measurement of binaries rates etc.) if the former dominate or provide a new cosmological probe if the result is robust to the unknown astrophysics. We observe that, since the angular power spectrum of the AGWB energy density has a frequency dependence, the relative importance of cosmological and astrophysical effects could  depend on the frequency band chosen, hence offering the possibility to distinguish different astrophysical processes. The cross-correlation with lensing and galaxy distribution opens a window on the understanding of the distribution of GW sources, in analogy e.g. with Ref.~\cite{Raccanelli:2016cud} albeit in a different context. 

Our treatment assumed that all the GW sources are located inside galaxies. This can easily be generalized to include more exotic scenarios, e.g. models in which dark matter is made of primordial back holes with mass of the order of 20-100 solar masses. Cross-corellations can indicate to which extent black-hole mergers are correlated with galaxies and hence study this hypothesis.

To conclude, we stress that in order to get quantitative predictions for the angular power spectrum of the AGWB energy density, the local physics processes generating GW have to be parametrized in detail. A numerical study of  the AGWB anisotropies, specifying local physics models, will be detailed  in our companion article \cite{Cusinprep}. 

%%%%%%%%%%%%%%%%%%%%%%%%%%%%%
\section*{Acknowledgements} GC thanks IAP for warm hospitality during this work. We thank Chiara Caprini,  Irina Dvorkin, Guillaume Faye,  Elisabeth Vangioni for discussions and comments on the early version of this text. We thank Enrico Barausse, Carlo Contaldi, Ruth Durrer, Pierre Fleury and Atsushi Nishizawa for their comments on the first version of the paper. The work of GC is supported by the Swiss National Fond Foundation.  The work of CP and JPU is made in the ILP LABEX (under reference ANR-10-LABX-63) was supported by French state funds managed by the ANR within the Investissements d'Avenir programme under reference ANR-11-IDEX-0004-02. 
%%%%%%%%%%%%%%%%%%%%%%%%%%%%%

\appendix 

\section{Reference frames}\label{Lorentz}

\begin{figure*}
\includegraphics[scale=0.58]{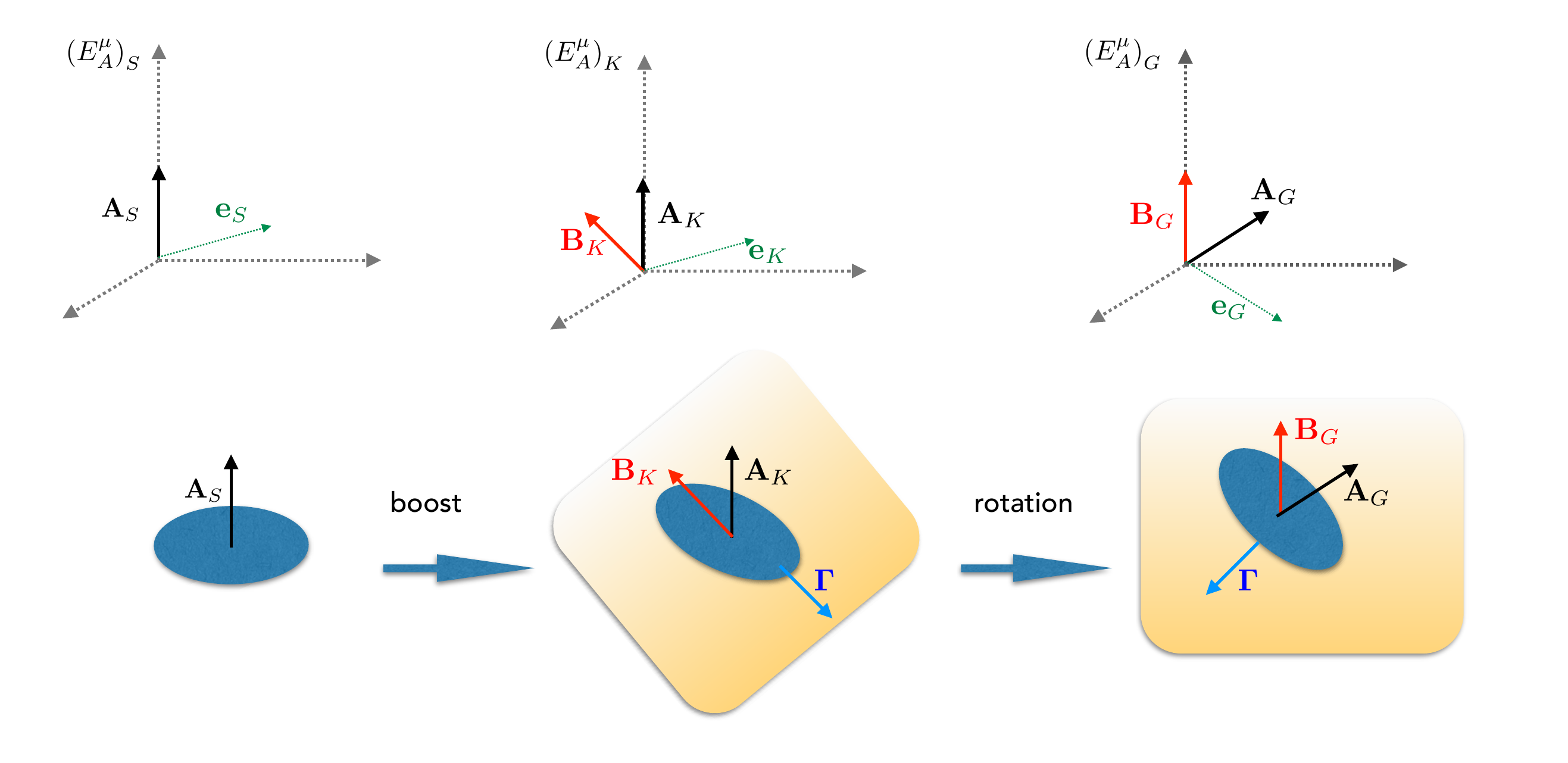}
\caption{\label{fig2}\textcolor{black}{Schematic representation of the change of reference frame we are considering. In the second line we represent in blue a generic source of GW and in yellow the galaxy in which the source is contained. }}
\end{figure*}

From now on we focus on a single source $i$ emitting GW in a galaxy. To simplify the notation, we omit the superscript $(i)$. We introduce the following reference frames
\begin{enumerate}
\item $S$: \,$\left(E_A^{\mu}\right)_{\Sou}\equiv (u^{\mu}, E_i^{\mu})_{\Sou}$ reference frame of the source, with the $z$ axis parallel to the axis of the source $\bA_{\Sou}$.
\item $K$:\, $\left(E_A^{\mu}\right)_{\Ka}\equiv (u^{\mu}, E_i^{\mu})_{\Ka}$ reference frame of the galaxy, with the $z$ axis parallel to the boosted axis of the source $\bA_{\Ka}$.  It is related to the first one by a boost with velocity $\bGamma$. 
\item $G$: \,$\left(E_A^{\mu}\right)_{\Gal}\equiv (u^{\mu}, E_i^{\mu})_{\Gal}$ reference frame of the galaxy, with the $z$ axis aligned with the axis of the galaxy $\bbB_{\Gal}$. It is related to the reference frame $\left(E_A^{\mu}\right)_{\Ka}$  through a rotation with Euler angles $\alpha, \beta, \gamma$. 
\end{enumerate}
We write explicitly the relation among the various frames. $K$ and $G$ are related by a rotation 
\begin{align}\label{rot}
(E_i)_{\Ka}&=R^i_j(\alpha, \beta, \gamma) (E^j)_{\Gal}\nn\\
&=R_{E_z^{\Gal}}(\gamma)^i_j R_{E_y^{\Gal}}(\beta)^j_k R_{E_z^{\Gal}}(\alpha)^k_{\ell} (E^{\ell})_{\Gal} \,.
\end{align}
$S$ and $K$ are related by a boost with velocity $\bGamma$. Explicitly 
\begin{align}
(E_{\mu}^{\rm A})_{\Ka}&=\Lambda^{\rm A}_{\,\,{\rm B}}(E_{\mu}^{{\rm B}})_{\Sou}\,,\\
(E^{\mu}_{\rm A})_{\Ka}&=(E^{\mu}_{\rm A})_{\Sou} \left(\Lambda^{-1}\right)^{\rm B}_{\,\,{\rm A}}=\Lambda_{\rm A}^{\,\,{\rm B}} (E^{\mu}_B)_{\Sou}\,,
\end{align}
with 
\begin{align}\label{Lambda}
&{\Lambda^0}_0 = \gamma \,,\,\quad {\Lambda^0}_i = {\Lambda^i}_0 = -
\gamma \Gamma_i\,,\nn\\
& {\Lambda^i}_j = \delta^i_j +
\frac{\gamma^2}{1+\gamma}\Gamma^i \Gamma_j\,,
\end{align}
and $\gamma^{-2}=1-\Gamma_i \Gamma^i$ and $\beta^2 \equiv \Gamma_i \Gamma^i$. A schematic representation of these reference frames is presented in Fig. \ref{fig2}.

\subsection{Situation in $S$ }

The energy spectrum emitted in the source reference frame $S$ depends on the angle between the line of sight and the axis of the source, i.e. $\bee_{\Sou}\cdot \bA_{\Sou}\equiv \mu_{\Sou}$, on the frequency $\nu_{\Sou}$, on the usual set of parameters $\theta_{\Gal}$ and on the parameters $\theta^{(i)}$ characterizing the source under exam. To simplify the notation, we omit in the following this last dependence. We therefore have 
\be\label{1}
\frac{\dd^3E_{\Sou}}{\dd\nu_{\Sou} \dd^2\Omega_{\Sou}}(\nu_{\Sou}, \bee_{\Sou}\cdot \bA_{\Sou}, \theta_{\Gal})=\frac{1}{2\pi}\frac{\dd^2E_{\Sou}}{\dd\nu_{\Sou} \dd\mu_{\Sou}}(\nu_{\Sou}, \mu_{\Sou}, \theta_{\Gal})\,.
\ee

\subsection{Situation in $K$ }
We need to boost the previous spectrum defined in $S$. The 4-vector $k^{\mu}$ of a graviton is decomposed in the two frames $S$ and $K$ as
\begin{align}
k^{\mu}&=\nu_{\Sou}\left(u_{\Sou}^{\mu}-e^{\mu}_{\Sou}\right)=\nu_{\Ka}\left(u_{\Ka}^{\mu}-e^{\mu}_{\Ka}\right)\,.
\end{align}
The relation between the frequency and the line of sight observed in the two frames is derived in Appendix G of  Ref.~\cite{Cusin:2016kqx}. At linear order in the boost velocity $\bGamma$,
\begin{align}\label{nnu}
\nu_{\Sou}&=\left(1\textcolor{black}{+}\bGamma\cdot \bee_{\Ka}\right)\nu_{\Ka}\,,\\
e_{\Sou}^i&=\left(1\textcolor{black}{-}\bGamma \cdot \bee_{\Ka}\right) e_{\Ka}^i\textcolor{black}{+}\Gamma^i\,,
\end{align}
where $e^i_{\Sou}\equiv E^i_{\mu} e^{\mu}_{\Sou}$ and $e^i_{\Ka}\equiv E^i_{\mu} e^{\mu}_{\Ka}$.  Analogously, introducing a vector $A^{\mu}$ defining the axis of the source, we introduce the decomposition
\begin{align}
A^{\mu}&=A_{\Sou}\left(u_{\Sou}^{\mu}+A^{\mu}_{\Sou}\right)=A_{\Ka}\left(u_{\Ka}^{\mu}+A^{\mu}_{\Ka}\right)\,.
\end{align}
We find up to linear order in $\bGamma$
\begin{align}
A_{\Sou}&=\left(1\textcolor{black}{-}\bGamma\cdot \bA_{\Ka}\right)A_{\Ka}\,,\\
A_{\Sou}^i&=\left(1\textcolor{black}{+}\bGamma \cdot \bA_{\Ka}\right) \bA_{\Ka}^i\textcolor{black}{-}\Gamma^i\,,
\end{align}
where $A^i_{\Sou}\equiv E^i_{\mu} A^{\mu}_{\Sou}$ and $A^i_{\Ka}\equiv E^i_{\mu} A^{\mu}_{\Ka}$. Defining $\mu_{\Sou}\equiv \cos \alpha_{\Sou}=\bA_{\Sou}\cdot \bee_{\Sou}$ and $\mu_{\Ka}\equiv \cos \alpha_{\Ka}=\bA_{\Ka}\cdot \bee_{\Ka}$, we get at linear order in $\bGamma$
\be\label{mmu}
\mu_{\Sou}=\mu_{\Ka}\textcolor{black}{-}(1+\mu_{\Ka})(\bee_{\Ka}-\bA_{\Ka})\cdot \bGamma\,.
\ee
This expression relates the angle identifying the orientation of a source in the source rest frame to the angles defining the orientation of the source in the galaxy rest frame.  If the source spectrum in the source rest frame has a rotational symmetry such that it depends only on $\mu_{\Sou}$, then this is not the case in the galaxy frame due to the aberration of the boost relating both frames, as can be seen by the dependence in $\bee_{\Ka}\cdot \bGamma$.
Indeed, the spherical coordinates of $\bee_{\Ka}$ in the frame K are obtained from
\be
e_{\Ka}^i=(E^i_\mu)_{\Ka} e_{\Ka}^\mu= (\cos \alpha_{\Ka} \cos \beta_{\Ka},\cos \alpha_{\Ka} \sin \beta_{\Ka},\sin \alpha_{\Ka})\,.
\ee

\subsection{Situation in $G$ }

We need to write the axis of the source in the reference frame G, using  (\ref{rot}). We get
\be
A^i_{\Gal}\equiv R^i_j(0, \beta, \gamma)A_{\Ka}^j\,,
\ee
\be
\bA_{\Gal}=\left(\sin\beta\cos\gamma\,, \sin\beta\sin\gamma\,, \cos\beta\right)\,.
\ee
The line of sight components will transform with (\ref{rot}). Explicitly 
\begin{align}
e_{\Gal}^i \equiv (E^i_\mu)_{\Gal} e_{\Gal}^\mu &= (\cos \alpha_{\Gal} \cos \beta_{\Gal},\cos \alpha_{\Gal} \sin \beta_{\Gal},\sin \alpha_{\Gal})\nonumber\\
&= R^i_j(0, \beta, \gamma)e_{\Ka}^j\,.
\end{align}
We now recall the relation between the reference frames $G$ and $K$ and we make use of  (\ref{mmu}), to relate the angle between the axis of the source and the line of sight in the source frame to the spherical angles defining the orientation of the source, the line of sight and the boost velocity in the galaxy frame. We indicate these angles and the magnitude of the boost velocity collectively as $\lambda_{\Gal}$, with $\lambda_{\Gal}\equiv (\bA_{\Gal}, \bGamma, \bee_{\Gal})$  and the resulting relation is of the type $\mu_{\Sou}(\lambda_{\Gal})$. We then relate the spectrum in the frame $G$ to the one in the source frame thanks to
\begin{align}\label{222}
&\frac{\dd^3 E_{\Gal}}{\dd\nu_{\Gal} \dd^2\Omega_{\Gal}}(\nu_{\Gal}, \lambda_{\Gal},   \theta_{\Gal}) = \left|\frac{\dd^2\Omega_{\Sou}}{\dd^2\Omega_{\Gal}}\right|\frac{\dd^3 E_{\Sou}}{\dd\nu_{\Sou} \dd^2\Omega_{\Sou}}(\nu_{\Sou}, \mu_{\Sou},\theta_{\Gal}) \,.
\end{align}

\subsection{Spectrum transformation}

The proper time of the observer in $K$ (and hence in $G$) is related to the proper time in $S$ by
\be
\dd t_{\Ka}=\frac{1}{\gamma} \dd t_{\Sou}=\dd t_{\Sou}+\mathcal{O}(\bGamma^2)\,.
\ee
It follows that at linear order in $\bGamma$ the two times can be identified.

The spectrum (\ref{222}) in the galaxy rest frame can be written in terms of quantities defined in the source rest frame as 
\begin{align}
&\frac{\dd^3 E_{\Gal}}{\dd^2\Omega_{\Gal} \dd\nu_{\Gal}}(\nu_{\Gal}, \lambda_{\Gal}, \theta_{\Gal})\\
& \quad=\frac{1}{2\pi}\frac{\dd^2 E_{\Sou}}{\dd\mu_{\Sou} \dd\nu_{\Sou}}(\nu_{\textcolor{black}{\Sou}}, \mu_{\Sou}, \theta_{\Gal})[1+\mathcal{O}(\bGamma)]\,,\nn
\end{align}
where we have used that the energy spectrum transforms as a scalar under boosts and substituted Eq. (\ref{mmu}) to assess that $\dd^2\Omega_{\Ka}/(2\pi \dd\mu_{\Sou})$ is unity up to linear terms in $\bGamma$.

%TODO. I let you modify from here, because it was me I would say it in words jumping directly to the sentence you write after these equations.

We insert these results in Eqs.~(\ref{LI}), (\ref{LM}) and make use of Eq. (\ref{0}).  By doing this, we get expressions for the effective luminosity of a galaxy in terms of local quantities defined in the source local frame. 
%\begin{eqnarray}
%\hspace{-2 em}\mathcal{L}_{\Gal}^I(\nu_{\Gal}, \theta_{\Gal})&=&\sum_{(i)}^I\int \dd\theta^{(i)} \mathcal{N}^{(i)}(\theta^{(i)}, \theta_{\Gal})\times \label{F1a}\\
%&& \hspace{-9 em}\times\int \frac{\dd\mu^{(i)}_{\Gal} }{4\pi}d\bA\Gal^{(i)} \frac{\dd E_{\Gal}^{(i)}}{\dd\mu_{\Gal}^{(i)}\dd\nu_{\Gal} \dd t_{\Gal}}(\nu_{\Gal},  \mu^{(i)}_{\Gal}, \theta_{\Gal})\,,\nn\\
%\mathcal{L}_{\Gal}^{M, SN}(\nu_{\Gal}, \theta_{\Gal})&=&\sum_{(i)}^{M, SN}\int \dd\theta^{(i)} \frac{d\mathcal{N}^{(i)}(\theta^{(i)}, \theta_{\Gal})}{\dd t_{\Gal}}\times  \label{F2a}\nn\\
%&& \hspace{-9 em}\times\int \frac{\dd\mu^{(i)}_{\Gal}}{4\pi} d\bA\Gal^{(i)} \frac{\dd E_{\Gal}^{(i)}}{\dd\mu_{\Gal}^{(i)}\dd\nu_{\Gal}}(\nu_{\Gal},  \mu^{(i)}_{\Gal}, \theta_{\Gal})\, ,
%\end{eqnarray}
Since linear contributions in $\bGamma$ vanish when averaging over $\bGamma$ with the distribution function $f(\theta_{\Gal}, \bGamma)$, in the final result for the effective luminosity Eqs.~(\ref{LI}), (\ref{LM}) energy, frequency and time in the source and galaxy frame can be simply identified, i.e. $\nu_{\Gal}=\nu_{\Sou}$, $E_{\Gal}=E_{\Sou}$  and $t_{\Gal}=t_{\Sou}$,  and the energy spectrum depends on directions only through $\mu_{\Gal}=\mu_{\Sou}$ (up to linear order in $\bGamma$). We find, explicitly up to first order in $\bGamma$ 
\begin{eqnarray}\label{LIA}
\mathcal{L}_{\Gal}^I\left(\nu_{\Gal}, \theta_{\Gal}\right)&\equiv& \sum_{(i)}^I\int \dd\theta^{(i)} \mathcal{N}^{(i)}(\theta^{(i)}, \theta_{\Gal})\nn\\
&& \hspace{-6 em} \times 
\int d\mu_{\Sou} \frac{\dd^3 E^{(i)}_{\Sou}}{\dd \mu_{\Sou}\dd t_{\Sou} d\nu_{\Sou}}(\nu_{\Sou}, \mu_{\Sou}, \theta^{(i)}, \theta_{\Gal})\,,
\end{eqnarray}
and 
\begin{align}\label{LMA}
&\mathcal{L}_{\Gal}^M(\nu_{\Gal}, \theta_{\Gal})=\nn\\
&\sum_{(i)}^M\int \dd\theta^{(i)}\, \frac{\dd\mathcal{N}^{(i)}}{\dd t_{\Sou}}(\theta^{(i)}, \theta_{\Gal})\nn\\
&\times \int \dd \mu_{\Sou} \frac{\dd^2 E_{\Sou}^{(i)}}{\dd\mu_{\Sou} \dd\nu_{\Sou}}(\nu_{\Sou}, \mu_{\Sou}, \theta^{(i)}, \theta_{\Gal})\,.
\end{align}

Summarizing, to get how the energy spectrum and the power emitted by a GW source transform under a boost with velocity $\bGamma$, three effects have to be considered: Doppler shift of frequencies, aberration effect and transformation of time. We have computed these effects and we have showed that up to linear order in $\bGamma$, we can identify the energy spectrum and the power in the reference frames of the source and of the galaxy since linear terms in $\bGamma$ coming from the Lorentz transformation cancel once averaged with the velocity distribution. We emphasize that no other effect comes into play.  
%In particular, since we do not resolve sources emitting GW inside a galaxy, it does not make sense to talk about relativistic beaming. A relativistic beaming effect, is due to the change of solid angle around the line of sight when considering two reference frames in relative motion. At the galaxy scale we do not resolve single sources, and the angular dependence of the emitted spectrum is due to the fact that we consider the possibility of an anisotropic emission (this angle is not an observable quantity, and we average over i
On the largest scale of our problem (cosmological scale), galaxies are considered as point-like sources and at this scale and the effect of a relativistic beaming is hidden in the distance-duality relation, see e.g. section 3.2.4. of \cite{Fleury:2015hgz}. 

\newpage
\bibliographystyle{utcaps}
\bibliography{BH+BH-refs.bib}

\providecommand{\href}[2]{#2}\begingroup\raggedright\begin{thebibliography}{10}

\bibitem{Abbott:2016blz}
{\bfseries Virgo, LIGO Scientific} Collaboration, B.~P. Abbott {\em et~al.},
  ``{Observation of Gravitational Waves from a Binary Black Hole Merger},''
  \href{http://dx.doi.org/10.1103/PhysRevLett.116.061102}{{\em Phys. Rev.
  Lett.} {\bfseries 116} no.~6, (2016) 061102},
\href{http://arxiv.org/abs/1602.03837}{{\ttfamily arXiv:1602.03837 [gr-qc]}}.
%%CITATION = ARXIV:1602.03837;%%.

\bibitem{Abbott:2016nmj}
{\bfseries Virgo, LIGO Scientific} Collaboration, B.~P. Abbott {\em et~al.},
  ``{GW151226: Observation of Gravitational Waves from a 22-Solar-Mass Binary
  Black Hole Coalescence},''
  \href{http://dx.doi.org/10.1103/PhysRevLett.116.241103}{{\em Phys. Rev.
  Lett.} {\bfseries 116} no.~24, (2016) 241103},
\href{http://arxiv.org/abs/1606.04855}{{\ttfamily arXiv:1606.04855 [gr-qc]}}.
%%CITATION = ARXIV:1606.04855;%%.

\bibitem{Abbott:2017vtc}
{\bfseries VIRGO, LIGO Scientific} Collaboration, B.~P. Abbott {\em et~al.},
  ``{GW170104: Observation of a 50-Solar-Mass Binary Black Hole Coalescence at
  Redshift 0.2},'' \href{http://dx.doi.org/10.1103/PhysRevLett.118.221101}{{\em
  Phys. Rev. Lett.} {\bfseries 118} no.~22, (2017) 221101},
\href{http://arxiv.org/abs/1706.01812}{{\ttfamily arXiv:1706.01812 [gr-qc]}}.
%%CITATION = ARXIV:1706.01812;%%.

\bibitem{Abbott:2017oio}
{\bfseries Virgo, LIGO Scientific} Collaboration, B.~P. Abbott {\em et~al.},
  ``{GW170814: A three-detector observation of gravitational waves from a
  binary black hole coalescence},'' {\em Submitted to: Phys. Rev. Lett.} (2017)
  ,
\href{http://arxiv.org/abs/1709.09660}{{\ttfamily arXiv:1709.09660 [gr-qc]}}.
%%CITATION = ARXIV:1709.09660;%%.

\bibitem{TheLIGOScientific:2016dpb}
{\bfseries Virgo, LIGO Scientific} Collaboration, B.~P. Abbott {\em et~al.},
  ``{Upper Limits on the Stochastic Gravitational-Wave Background from Advanced
  LIGO First Observing Run},''
  \href{http://dx.doi.org/10.1103/PhysRevLett.118.121101,
  10.1103/PhysRevLett.119.029901}{{\em Phys. Rev. Lett.} {\bfseries 118}
  no.~12, (2017) 121101}, \href{http://arxiv.org/abs/1612.02029}{{\ttfamily
  arXiv:1612.02029 [gr-qc]}}.
[Erratum: Phys. Rev. Lett.119,no.2,029901(2017)].
%%CITATION = ARXIV:1612.02029;%%.

\bibitem{Aasi:2014zwg}
{\bfseries VIRGO, LIGO Scientific} Collaboration, J.~Aasi {\em et~al.},
  ``{Improved Upper Limits on the Stochastic Gravitational-Wave Background from
  2009-2010 LIGO and Virgo Data},''
  \href{http://dx.doi.org/10.1103/PhysRevLett.113.231101}{{\em Phys. Rev.
  Lett.} {\bfseries 113} no.~23, (2014) 231101},
\href{http://arxiv.org/abs/1406.4556}{{\ttfamily arXiv:1406.4556 [gr-qc]}}.
%%CITATION = ARXIV:1406.4556;%%.

\bibitem{Maggiore:1999vm}
M.~Maggiore, ``{Gravitational wave experiments and early universe cosmology},''
  \href{http://dx.doi.org/10.1016/S0370-1573(99)00102-7}{{\em Phys. Rept.}
  {\bfseries 331} (2000) 283--367},
\href{http://arxiv.org/abs/gr-qc/9909001}{{\ttfamily arXiv:gr-qc/9909001
  [gr-qc]}}.
%%CITATION = GR-QC/9909001;%%.

\bibitem{Allen:1996vm}
B.~Allen, ``{The Stochastic gravity wave background: Sources and detection},''
  in {\em {Relativistic gravitation and gravitational radiation. Proceedings,
  School of Physics, Les Houches, France, September 26-October 6, 1995}},
  pp.~373--417.
\newblock 1996.
\newblock \href{http://arxiv.org/abs/gr-qc/9604033}{{\ttfamily
  arXiv:gr-qc/9604033 [gr-qc]}}.
\newblock
\url{http://alice.cern.ch/format/showfull?sysnb=0223102}.
\newblock
%%CITATION = GR-QC/9604033;%%.

\bibitem{Smith:2006nka}
T.~L. Smith, E.~Pierpaoli, and M.~Kamionkowski, ``{A new cosmic microwave
  background constraint to primordial gravitational waves},''
  \href{http://dx.doi.org/10.1103/PhysRevLett.97.021301}{{\em Phys. Rev. Lett.}
  {\bfseries 97} (2006) 021301},
\href{http://arxiv.org/abs/astro-ph/0603144}{{\ttfamily arXiv:astro-ph/0603144
  [astro-ph]}}.
%%CITATION = ASTRO-PH/0603144;%%.

\bibitem{Henrot-Versille:2014jua}
S.~Henrot-Versille {\em et~al.}, ``{Improved constraint on the primordial
  gravitational-wave density using recent cosmological data and its impact on
  cosmic string models},''
  \href{http://dx.doi.org/10.1088/0264-9381/32/4/045003}{{\em Class. Quant.
  Grav.} {\bfseries 32} no.~4, (2015) 045003},
\href{http://arxiv.org/abs/1408.5299}{{\ttfamily arXiv:1408.5299
  [astro-ph.CO]}}.
%%CITATION = ARXIV:1408.5299;%%.

\bibitem{Shannon:2013wma}
R.~M. Shannon {\em et~al.}, ``{Gravitational-wave Limits from Pulsar Timing
  Constrain Supermassive Black Hole Evolution},''
  \href{http://dx.doi.org/10.1126/science.1238012}{{\em Science} {\bfseries
  342} no.~6156, (2013) 334--337},
\href{http://arxiv.org/abs/1310.4569}{{\ttfamily arXiv:1310.4569
  [astro-ph.CO]}}.
%%CITATION = ARXIV:1310.4569;%%.

\bibitem{Allen:1996gp}
B.~Allen and A.~C. Ottewill, ``{Detection of anisotropies in the gravitational
  wave stochastic background},''
  \href{http://dx.doi.org/10.1103/PhysRevD.56.545}{{\em Phys. Rev.} {\bfseries
  D56} (1997) 545--563},
\href{http://arxiv.org/abs/gr-qc/9607068}{{\ttfamily arXiv:gr-qc/9607068
  [gr-qc]}}.
%%CITATION = GR-QC/9607068;%%.

\bibitem{Cornish:2001hg}
N.~J. Cornish, ``{Mapping the gravitational wave background},''
  \href{http://dx.doi.org/10.1088/0264-9381/18/20/307}{{\em Class. Quant.
  Grav.} {\bfseries 18} (2001) 4277--4292},
\href{http://arxiv.org/abs/astro-ph/0105374}{{\ttfamily arXiv:astro-ph/0105374
  [astro-ph]}}.
%%CITATION = ASTRO-PH/0105374;%%.

\bibitem{Mitra:2007mc}
S.~Mitra, S.~Dhurandhar, T.~Souradeep, A.~Lazzarini, V.~Mandic, S.~Bose, and
  S.~Ballmer, ``{Gravitational wave radiometry: Mapping a stochastic
  gravitational wave background},''
  \href{http://dx.doi.org/10.1103/PhysRevD.77.042002}{{\em Phys. Rev.}
  {\bfseries D77} (2008) 042002},
\href{http://arxiv.org/abs/0708.2728}{{\ttfamily arXiv:0708.2728 [gr-qc]}}.
%%CITATION = ARXIV:0708.2728;%%.

\bibitem{Thrane:2009fp}
E.~Thrane, S.~Ballmer, J.~D. Romano, S.~Mitra, D.~Talukder, S.~Bose, and
  V.~Mandic, ``{Probing the anisotropies of a stochastic gravitational-wave
  background using a network of ground-based laser interferometers},''
  \href{http://dx.doi.org/10.1103/PhysRevD.80.122002}{{\em Phys. Rev.}
  {\bfseries D80} (2009) 122002},
\href{http://arxiv.org/abs/0910.0858}{{\ttfamily arXiv:0910.0858
  [astro-ph.IM]}}.
%%CITATION = ARXIV:0910.0858;%%.

\bibitem{Romano:2015uma}
J.~D. Romano, S.~R. Taylor, N.~J. Cornish, J.~Gair, C.~M.~F. Mingarelli, and
  R.~van Haasteren, ``{Phase-coherent mapping of gravitational-wave backgrounds
  using ground-based laser interferometers},''
  \href{http://dx.doi.org/10.1103/PhysRevD.92.042003}{{\em Phys. Rev.}
  {\bfseries D92} no.~4, (2015) 042003},
\href{http://arxiv.org/abs/1505.07179}{{\ttfamily arXiv:1505.07179 [gr-qc]}}.
%%CITATION = ARXIV:1505.07179;%%.

\bibitem{Romano:2016dpx}
J.~D. Romano and N.~J. Cornish, ``{Detection methods for stochastic
  gravitational-wave backgrounds: A unified treatment},''
\href{http://arxiv.org/abs/1608.06889}{{\ttfamily arXiv:1608.06889 [gr-qc]}}.
%%CITATION = ARXIV:1608.06889;%%.

\bibitem{TheLIGOScientific:2016xzw}
{\bfseries Virgo, LIGO Scientific} Collaboration, B.~P. Abbott {\em et~al.},
  ``{Directional Limits on Persistent Gravitational Waves from Advanced LIGO?s
  First Observing Run},''
  \href{http://dx.doi.org/10.1103/PhysRevLett.118.121102}{{\em Phys. Rev.
  Lett.} {\bfseries 118} no.~12, (2017) 121102},
\href{http://arxiv.org/abs/1612.02030}{{\ttfamily arXiv:1612.02030 [gr-qc]}}.
%%CITATION = ARXIV:1612.02030;%%.

\bibitem{Mingarelli:2013dsa}
C.~M.~F. Mingarelli, T.~Sidery, I.~Mandel, and A.~Vecchio, ``{Characterizing
  gravitational wave stochastic background anisotropy with pulsar timing
  arrays},'' \href{http://dx.doi.org/10.1103/PhysRevD.88.062005}{{\em Phys.
  Rev.} {\bfseries D88} no.~6, (2013) 062005},
\href{http://arxiv.org/abs/1306.5394}{{\ttfamily arXiv:1306.5394
  [astro-ph.HE]}}.
%%CITATION = ARXIV:1306.5394;%%.

\bibitem{Taylor:2013esa}
S.~R. Taylor and J.~R. Gair, ``{Searching For Anisotropic Gravitational-wave
  Backgrounds Using Pulsar Timing Arrays},''
  \href{http://dx.doi.org/10.1103/PhysRevD.88.084001}{{\em Phys. Rev.}
  {\bfseries D88} (2013) 084001},
\href{http://arxiv.org/abs/1306.5395}{{\ttfamily arXiv:1306.5395 [gr-qc]}}.
%%CITATION = ARXIV:1306.5395;%%.

\bibitem{Gair:2014rwa}
J.~Gair, J.~D. Romano, S.~Taylor, and C.~M.~F. Mingarelli, ``{Mapping
  gravitational-wave backgrounds using methods from CMB analysis: Application
  to pulsar timing arrays},''
  \href{http://dx.doi.org/10.1103/PhysRevD.90.082001}{{\em Phys. Rev.}
  {\bfseries D90} no.~8, (2014) 082001},
\href{http://arxiv.org/abs/1406.4664}{{\ttfamily arXiv:1406.4664 [gr-qc]}}.
%%CITATION = ARXIV:1406.4664;%%.

\bibitem{Regimbau:2011rp}
T.~Regimbau, ``{The astrophysical gravitational wave stochastic background},''
  \href{http://dx.doi.org/10.1088/1674-4527/11/4/001}{{\em Res. Astron.
  Astrophys.} {\bfseries 11} (2011) 369--390},
\href{http://arxiv.org/abs/1101.2762}{{\ttfamily arXiv:1101.2762
  [astro-ph.CO]}}.
%%CITATION = ARXIV:1101.2762;%%.

\bibitem{PeterUzan2005}
P.~Peter and J.-P. Uzan, {\em {Primordial Cosmology}}.
\newblock Oxford Graduate Texts. Oxford University Press, 2005.
\newblock
  \url{https://global.oup.com/academic/product/primordial-cosmology-9780199665150?cc=fr&lang=en&}.

\bibitem{Dufaux:2007pt}
J.~F. Dufaux, A.~Bergman, G.~N. Felder, L.~Kofman, and J.-P. Uzan, ``{Theory
  and Numerics of Gravitational Waves from Preheating after Inflation},''
  \href{http://dx.doi.org/10.1103/PhysRevD.76.123517}{{\em Phys. Rev.}
  {\bfseries D76} (2007) 123517},
\href{http://arxiv.org/abs/0707.0875}{{\ttfamily arXiv:0707.0875 [astro-ph]}}.
%%CITATION = ARXIV:0707.0875;%%.

\bibitem{Vilenkin:1981bx}
A.~Vilenkin, ``{Gravitational radiation from cosmic strings},''
\href{http://dx.doi.org/10.1016/0370-2693(81)91144-8}{{\em Phys. Lett.}
  {\bfseries B107} (1981) 47--50}.
%%CITATION = PHLTA,B107,47;%%.

\bibitem{Hogan:1984is}
C.~J. Hogan and M.~J. Rees, ``{Gravitational interactions of cosmic strings},''
\href{http://dx.doi.org/10.1038/311109a0}{{\em Nature} {\bfseries 311} (1984)
  109--113}.
%%CITATION = NATUA,311,109;%%.

\bibitem{Vachaspati:1984gt}
T.~Vachaspati and A.~Vilenkin, ``{Gravitational Radiation from Cosmic
  Strings},''
\href{http://dx.doi.org/10.1103/PhysRevD.31.3052}{{\em Phys. Rev.} {\bfseries
  D31} (1985) 3052}.
%%CITATION = PHRVA,D31,3052;%%.

\bibitem{Caldwell:1991jj}
R.~R. Caldwell and B.~Allen, ``{Cosmological constraints on cosmic string
  gravitational radiation},''
\href{http://dx.doi.org/10.1103/PhysRevD.45.3447}{{\em Phys. Rev.} {\bfseries
  D45} (1992) 3447--3468}.
%%CITATION = PHRVA,D45,3447;%%.

\bibitem{Kuroyanagi:2016ugi}
S.~Kuroyanagi, K.~Takahashi, N.~Yonemaru, and H.~Kumamoto, ``{Anisotropies in
  the gravitational wave background as a probe of the cosmic string network},''
  \href{http://dx.doi.org/10.1103/PhysRevD.95.043531}{{\em Phys. Rev.}
  {\bfseries D95} no.~4, (2017) 043531},
\href{http://arxiv.org/abs/1604.00332}{{\ttfamily arXiv:1604.00332
  [astro-ph.CO]}}.
%%CITATION = ARXIV:1604.00332;%%.

\bibitem{Caprini:2009fx}
C.~Caprini, R.~Durrer, T.~Konstandin, and G.~Servant, ``{General Properties of
  the Gravitational Wave Spectrum from Phase Transitions},''
  \href{http://dx.doi.org/10.1103/PhysRevD.79.083519}{{\em Phys. Rev.}
  {\bfseries D79} (2009) 083519},
\href{http://arxiv.org/abs/0901.1661}{{\ttfamily arXiv:0901.1661
  [astro-ph.CO]}}.
%%CITATION = ARXIV:0901.1661;%%.

\bibitem{Caprini:2015zlo}
C.~Caprini {\em et~al.}, ``{Science with the space-based interferometer eLISA.
  II: Gravitational waves from cosmological phase transitions},''
  \href{http://dx.doi.org/10.1088/1475-7516/2016/04/001}{{\em JCAP} {\bfseries
  1604} no.~04, (2016) 001},
\href{http://arxiv.org/abs/1512.06239}{{\ttfamily arXiv:1512.06239
  [astro-ph.CO]}}.
%%CITATION = ARXIV:1512.06239;%%.

\bibitem{Caprini:2001nb}
C.~Caprini and R.~Durrer, ``{Gravitational wave production: A Strong constraint
  on primordial magnetic fields},''
  \href{http://dx.doi.org/10.1103/PhysRevD.65.023517}{{\em Phys. Rev.}
  {\bfseries D65} (2001) 023517},
\href{http://arxiv.org/abs/astro-ph/0106244}{{\ttfamily arXiv:astro-ph/0106244
  [astro-ph]}}.
%%CITATION = ASTRO-PH/0106244;%%.

\bibitem{Binetruy:2012ze}
P.~Binetruy, A.~Bohe, C.~Caprini, and J.-F. Dufaux, ``{Cosmological Backgrounds
  of Gravitational Waves and eLISA/NGO: Phase Transitions, Cosmic Strings and
  Other Sources},'' \href{http://dx.doi.org/10.1088/1475-7516/2012/06/027}{{\em
  JCAP} {\bfseries 1206} (2012) 027},
\href{http://arxiv.org/abs/1201.0983}{{\ttfamily arXiv:1201.0983 [gr-qc]}}.
%%CITATION = ARXIV:1201.0983;%%.

\bibitem{Buonanno:2014aza}
A.~Buonanno and B.~S. Sathyaprakash, ``{Sources of Gravitational Waves: Theory
  and Observations},'' \href{http://arxiv.org/abs/1410.7832}{{\ttfamily
  arXiv:1410.7832 [gr-qc]}}.
\url{https://inspirehep.net/record/1324934/files/arXiv:1410.7832.pdf}.
%%CITATION = ARXIV:1410.7832;%%.

\bibitem{TheLIGOScientific:2016wyq}
{\bfseries Virgo, LIGO Scientific} Collaboration, B.~P. Abbott {\em et~al.},
  ``{GW150914: Implications for the stochastic gravitational wave background
  from binary black holes},''
  \href{http://dx.doi.org/10.1103/PhysRevLett.116.131102}{{\em Phys. Rev.
  Lett.} {\bfseries 116} no.~13, (2016) 131102},
\href{http://arxiv.org/abs/1602.03847}{{\ttfamily arXiv:1602.03847 [gr-qc]}}.
%%CITATION = ARXIV:1602.03847;%%.

\bibitem{Regimbau:2016ike}
T.~Regimbau, M.~Evans, N.~Christensen, E.~Katsavounidis, B.~Sathyaprakash, and
  S.~Vitale, ``{Digging deeper: Observing primordial gravitational waves below
  the binary black hole produced stochastic background},''
\href{http://arxiv.org/abs/1611.08943}{{\ttfamily arXiv:1611.08943
  [astro-ph.CO]}}.
%%CITATION = ARXIV:1611.08943;%%.

\bibitem{Mandic:2016lcn}
V.~Mandic, S.~Bird, and I.~Cholis, ``{Stochastic Gravitational-Wave Background
  due to Primordial Binary Black Hole Mergers},''
  \href{http://dx.doi.org/10.1103/PhysRevLett.117.201102}{{\em Phys. Rev.
  Lett.} {\bfseries 117} no.~20, (2016) 201102},
\href{http://arxiv.org/abs/1608.06699}{{\ttfamily arXiv:1608.06699
  [astro-ph.CO]}}.
%%CITATION = ARXIV:1608.06699;%%.

\bibitem{Dvorkin:2016okx}
I.~Dvorkin, J.-P. Uzan, E.~Vangioni, and J.~Silk, ``{Synthetic model of the
  gravitational wave background from evolving binary compact objects},''
  \href{http://dx.doi.org/10.1103/PhysRevD.94.103011}{{\em Phys. Rev.}
  {\bfseries D94} no.~10, (2016) 103011},
\href{http://arxiv.org/abs/1607.06818}{{\ttfamily arXiv:1607.06818
  [astro-ph.HE]}}.
%%CITATION = ARXIV:1607.06818;%%.

\bibitem{Nakazato:2016nkj}
K.~Nakazato, Y.~Niino, and N.~Sago, ``{Gravitational-Wave Background from
  Binary Mergers and Metallicity Evolution of Galaxies},''
  \href{http://dx.doi.org/10.3847/0004-637X/832/2/146}{{\em Astrophys. J.}
  {\bfseries 832} no.~2, (2016) 146},
\href{http://arxiv.org/abs/1605.02146}{{\ttfamily arXiv:1605.02146
  [astro-ph.HE]}}.
%%CITATION = ARXIV:1605.02146;%%.

\bibitem{Dvorkin:2016wac}
I.~Dvorkin, E.~Vangioni, J.~Silk, J.-P. Uzan, and K.~A. Olive,
  ``{Metallicity-constrained merger rates of binary black holes and the
  stochastic gravitational wave background},''
  \href{http://dx.doi.org/10.1093/mnras/stw1477}{{\em Mon. Not. Roy. Astron.
  Soc.} {\bfseries 461} no.~4, (2016) 3877--3885},
\href{http://arxiv.org/abs/1604.04288}{{\ttfamily arXiv:1604.04288
  [astro-ph.HE]}}.
%%CITATION = ARXIV:1604.04288;%%.

\bibitem{Evangelista:2014oba}
E.~F.~D. Evangelista and J.~C.~N. Araujo, ``{The Gravitational Wave Background
  from Coalescing Compact Binaries: A New Method},''
  \href{http://dx.doi.org/10.1007/s13538-014-0272-0}{{\em Braz. J. Phys.}
  {\bfseries 44} no.~6, (2014) 824--831},
\href{http://arxiv.org/abs/1504.06605}{{\ttfamily arXiv:1504.06605
  [astro-ph.CO]}}.
%%CITATION = ARXIV:1504.06605;%%.

\bibitem{Kelley:2017lek}
L.~Z. Kelley, L.~Blecha, L.~Hernquist, and A.~Sesana, ``{The Gravitational Wave
  Background from Massive Black Hole Binaries in Illustris: spectral features
  and time to detection with pulsar timing arrays},''
\href{http://arxiv.org/abs/1702.02180}{{\ttfamily arXiv:1702.02180
  [astro-ph.HE]}}.
%%CITATION = ARXIV:1702.02180;%%.

\bibitem{Crocker:2015taa}
K.~Crocker, V.~Mandic, T.~Regimbau, K.~Belczynski, W.~Gladysz, K.~Olive,
  T.~Prestegard, and E.~Vangioni, ``{Model of the stochastic gravitational-wave
  background due to core collapse to black holes},''
  \href{http://dx.doi.org/10.1103/PhysRevD.92.063005}{{\em Phys. Rev.}
  {\bfseries D92} no.~6, (2015) 063005},
\href{http://arxiv.org/abs/1506.02631}{{\ttfamily arXiv:1506.02631 [gr-qc]}}.
%%CITATION = ARXIV:1506.02631;%%.

\bibitem{Surace:2015ppq}
M.~Surace, K.~D. Kokkotas, and P.~Pnigouras, ``{The stochastic background of
  gravitational waves due to the $f$-mode instability in neutron stars},''
  \href{http://dx.doi.org/10.1051/0004-6361/201527197}{{\em Astron. Astrophys.}
  {\bfseries 586} (2016) A86},
\href{http://arxiv.org/abs/1512.02502}{{\ttfamily arXiv:1512.02502
  [astro-ph.CO]}}.
%%CITATION = ARXIV:1512.02502;%%.

\bibitem{Talukder:2014eba}
D.~Talukder, E.~Thrane, S.~Bose, and T.~Regimbau, ``{Measuring neutron-star
  ellipticity with measurements of the stochastic gravitational-wave
  background},'' \href{http://dx.doi.org/10.1103/PhysRevD.89.123008}{{\em Phys.
  Rev.} {\bfseries D89} no.~12, (2014) 123008},
\href{http://arxiv.org/abs/1404.4025}{{\ttfamily arXiv:1404.4025 [gr-qc]}}.
%%CITATION = ARXIV:1404.4025;%%.

\bibitem{Lasky:2013jfa}
P.~D. Lasky, M.~F. Bennett, and A.~Melatos, ``{Stochastic gravitational wave
  background from hydrodynamic turbulence in differentially rotating neutron
  stars},'' \href{http://dx.doi.org/10.1103/PhysRevD.87.063004}{{\em Phys.
  Rev.} {\bfseries D87} no.~6, (2013) 063004},
\href{http://arxiv.org/abs/1302.6033}{{\ttfamily arXiv:1302.6033
  [astro-ph.HE]}}.
%%CITATION = ARXIV:1302.6033;%%.

\bibitem{Crocker:2017agi}
K.~Crocker, T.~Prestegard, V.~Mandic, T.~Regimbau, K.~Olive, and E.~Vangioni,
  ``{A Systematic Study of the Stochastic Gravitational-Wave Background due to
  Stellar Core Collapse},''
\href{http://arxiv.org/abs/1701.02638}{{\ttfamily arXiv:1701.02638
  [astro-ph.CO]}}.
%%CITATION = ARXIV:1701.02638;%%.

\bibitem{Kowalska:2012ba}
I.~Kowalska, T.~Bulik, and K.~Belczynski, ``{Gravitational wave background from
  population III binaries},''
  \href{http://dx.doi.org/10.1051/0004-6361/201118604}{{\em Astron. Astrophys.}
  {\bfseries 541} (2012) A120},
\href{http://arxiv.org/abs/1202.3346}{{\ttfamily arXiv:1202.3346
  [astro-ph.CO]}}.
%%CITATION = ARXIV:1202.3346;%%.

\bibitem{Moore:2014lga}
C.~J. Moore, R.~H. Cole, and C.~P.~L. Berry, ``{Gravitational-wave sensitivity
  curves},'' \href{http://dx.doi.org/10.1088/0264-9381/32/1/015014}{{\em Class.
  Quant. Grav.} {\bfseries 32} no.~1, (2015) 015014},
\href{http://arxiv.org/abs/1408.0740}{{\ttfamily arXiv:1408.0740 [gr-qc]}}.
%%CITATION = ARXIV:1408.0740;%%.

\bibitem{Evans:2016mbw}
{\bfseries LIGO Scientific} Collaboration, B.~P. Abbott {\em et~al.},
  ``{Exploring the Sensitivity of Next Generation Gravitational Wave
  Detectors},'' \href{http://dx.doi.org/10.1088/1361-6382/aa51f4}{{\em Class.
  Quant. Grav.} {\bfseries 34} no.~4, (2017) 044001},
\href{http://arxiv.org/abs/1607.08697}{{\ttfamily arXiv:1607.08697
  [astro-ph.IM]}}.
%%CITATION = ARXIV:1607.08697;%%.

\bibitem{Taylor2015}
S.~R. Taylor {\em et~al.}, ``{Limits on anisotropy in the nanohertz stochastic
  gravitational-wave background},''
  \href{http://dx.doi.org/10.1103/PhysRevLett.115.041101}{{\em Phys. Rev.
  Lett.} {\bfseries 115} no.~4, (2015) 041101},
\href{http://arxiv.org/abs/1506.08817}{{\ttfamily arXiv:1506.08817
  [astro-ph.HE]}}.
%%CITATION = ARXIV:1506.08817;%%.

\bibitem{Sesana2008}
A.~Sesana, A.~Vecchio, and C.~N. Colacino, ``{The stochastic gravitational-wave
  background from massive black hole binary systems: implications for
  observations with Pulsar Timing Arrays},''
  \href{http://dx.doi.org/10.1111/j.1365-2966.2008.13682.x}{{\em Mon. Not. Roy.
  Astron. Soc.} {\bfseries 390} (2008) 192},
\href{http://arxiv.org/abs/0804.4476}{{\ttfamily arXiv:0804.4476 [astro-ph]}}.
%%CITATION = ARXIV:0804.4476;%%.

\bibitem{Cusinprep}
G.~Cusin, J.-P. Uzan, and C.~Pitrou, ``{Anisotropy of the astrophysical
  gravitational waves background I:Predictions for the angular power spectrum
  (In preparation)},''.

\bibitem{Carr:2016drx}
B.~Carr, F.~Kuhnel, and M.~Sandstad, ``{Primordial Black Holes as Dark
  Matter},'' \href{http://dx.doi.org/10.1103/PhysRevD.94.083504}{{\em Phys.
  Rev.} {\bfseries D94} no.~8, (2016) 083504},
\href{http://arxiv.org/abs/1607.06077}{{\ttfamily arXiv:1607.06077
  [astro-ph.CO]}}.
%%CITATION = ARXIV:1607.06077;%%.

\bibitem{Bird:2016dcv}
S.~Bird, I.~Cholis, J.~B. Muñoz, Y.~Ali-Haïmoud, M.~Kamionkowski, E.~D. Kovetz,
  A.~Raccanelli, and A.~G. Riess, ``{Did LIGO detect dark matter?},''
  \href{http://dx.doi.org/10.1103/PhysRevLett.116.201301}{{\em Phys. Rev.
  Lett.} {\bfseries 116} no.~20, (2016) 201301},
\href{http://arxiv.org/abs/1603.00464}{{\ttfamily arXiv:1603.00464
  [astro-ph.CO]}}.
%%CITATION = ARXIV:1603.00464;%%.

\bibitem{Clesse:2016ajp}
S.~Clesse and J.~García-Bellido, ``{Detecting the gravitational wave background
  from primordial black hole dark matter},''
\href{http://arxiv.org/abs/1610.08479}{{\ttfamily arXiv:1610.08479
  [astro-ph.CO]}}.
%%CITATION = ARXIV:1610.08479;%%.

\bibitem{Raccanelli:2016cud}
A.~Raccanelli, E.~D. Kovetz, S.~Bird, I.~Cholis, and J.~B. Munoz,
  ``{Determining the progenitors of merging black-hole binaries},''
  \href{http://dx.doi.org/10.1103/PhysRevD.94.023516}{{\em Phys. Rev.}
  {\bfseries D94} no.~2, (2016) 023516},
\href{http://arxiv.org/abs/1605.01405}{{\ttfamily arXiv:1605.01405
  [astro-ph.CO]}}.
%%CITATION = ARXIV:1605.01405;%%.

\bibitem{Namikawa:2016edr}
T.~Namikawa, A.~Nishizawa, and A.~Taruya, ``{Detecting Black-Hole Binary
  Clustering via the Second-Generation Gravitational-Wave Detectors},''
  \href{http://dx.doi.org/10.1103/PhysRevD.94.024013}{{\em Phys. Rev.}
  {\bfseries D94} no.~2, (2016) 024013},
\href{http://arxiv.org/abs/1603.08072}{{\ttfamily arXiv:1603.08072
  [astro-ph.CO]}}.
%%CITATION = ARXIV:1603.08072;%%.

\bibitem{Contaldi:2016koz}
C.~R. Contaldi, ``{Anisotropies of Gravitational Wave Backgrounds: A Line Of
  Sight Approach},''
\href{http://arxiv.org/abs/1609.08168}{{\ttfamily arXiv:1609.08168
  [astro-ph.CO]}}.
%%CITATION = ARXIV:1609.08168;%%.

\bibitem{Namikawa:2015prh}
T.~Namikawa, A.~Nishizawa, and A.~Taruya, ``{Anisotropies of gravitational-wave
  standard sirens as a new cosmological probe without redshift information},''
  \href{http://dx.doi.org/10.1103/PhysRevLett.116.121302}{{\em Phys. Rev.
  Lett.} {\bfseries 116} no.~12, (2016) 121302},
\href{http://arxiv.org/abs/1511.04638}{{\ttfamily arXiv:1511.04638
  [astro-ph.CO]}}.
%%CITATION = ARXIV:1511.04638;%%.

\bibitem{Bertacca:2017vod}
D.~Bertacca, A.~Raccanelli, N.~Bartolo, and S.~Matarrese, ``{Cosmological
  perturbation effects on gravitational-wave luminosity distance estimates},''
\href{http://arxiv.org/abs/1702.01750}{{\ttfamily arXiv:1702.01750 [gr-qc]}}.
%%CITATION = ARXIV:1702.01750;%%.

\bibitem{Olmez:2011cg}
S.~Olmez, V.~Mandic, and X.~Siemens, ``{Anisotropies in the Gravitational-Wave
  Stochastic Background},''
  \href{http://dx.doi.org/10.1088/1475-7516/2012/07/009}{{\em JCAP} {\bfseries
  1207} (2012) 009},
\href{http://arxiv.org/abs/1106.5555}{{\ttfamily arXiv:1106.5555
  [astro-ph.CO]}}.
%%CITATION = ARXIV:1106.5555;%%.

\bibitem{Oguri:2016dgk}
M.~Oguri, ``{Measuring the distance-redshift relation with the
  cross-correlation of gravitational wave standard sirens and galaxies},''
  \href{http://dx.doi.org/10.1103/PhysRevD.93.083511}{{\em Phys. Rev.}
  {\bfseries D93} no.~8, (2016) 083511},
\href{http://arxiv.org/abs/1603.02356}{{\ttfamily arXiv:1603.02356
  [astro-ph.CO]}}.
%%CITATION = ARXIV:1603.02356;%%.

\bibitem{Uzan:2004my}
J.-P. Uzan, N.~Aghanim, and Y.~Mellier, ``{The Distance duality relation from
  x-ray and SZ observations of clusters},''
  \href{http://dx.doi.org/10.1103/PhysRevD.70.083533}{{\em Phys. Rev.}
  {\bfseries D70} (2004) 083533},
\href{http://arxiv.org/abs/astro-ph/0405620}{{\ttfamily arXiv:astro-ph/0405620
  [astro-ph]}}.
%%CITATION = ASTRO-PH/0405620;%%.

\bibitem{Etherington1933}
I.~Etherington, ``{LX. On the Definition of Distance in General Relativity},''
  {\em Philosophical Magazine} {\bfseries 15} (1933) 761--773.

\bibitem{Maggiore:1900zz}
M.~Maggiore, {\em {Gravitational Waves. Vol. 1: Theory and Experiments}}.
\newblock Oxford Master Series in Physics. Oxford University Press, 2007.
\newblock
\url{http://www.oup.com/uk/catalogue/?ci=9780198570745}.
\newblock
%%CITATION = INSPIRE-768483;%%.

\bibitem{DeruelleBook}
N.~Deruelle and J.-P. Uzan, {\em {Th\'eories de la relativit\'e}}.
\newblock Collection Echelles. Belin, 2014.
\newblock \url{https://www.belin-education.com/theories-de-la-relativite}.

\bibitem{1961RSPSA.264..309S}
R.~K. Sachs, ``{Gravitational waves in general relativity. 6. The outgoing
  radiation condition},''
\href{http://dx.doi.org/10.1098/rspa.1961.0202}{{\em Proc. Roy. Soc. Lond.}
  {\bfseries A264} (1961) 309--338}.
%%CITATION = PRSLA,A264,309;%%.

\bibitem{Isaacson:1967zz}
R.~A. Isaacson, ``{Gravitational Radiation in the Limit of High Frequency. I.
  The Linear Approximation and Geometrical Optics},''
\href{http://dx.doi.org/10.1103/PhysRev.166.1263}{{\em Phys. Rev.} {\bfseries
  166} (1967) 1263--1271}.
%%CITATION = PHRVA,166,1263;%%.

\bibitem{Bassett:2003vu}
B.~A. Bassett and M.~Kunz, ``{Cosmic distance-duality as a probe of exotic
  physics and acceleration},''
  \href{http://dx.doi.org/10.1103/PhysRevD.69.101305}{{\em Phys. Rev.}
  {\bfseries D69} (2004) 101305},
\href{http://arxiv.org/abs/astro-ph/0312443}{{\ttfamily arXiv:astro-ph/0312443
  [astro-ph]}}.
%%CITATION = ASTRO-PH/0312443;%%.

\bibitem{Ellis:2013cu}
G.~F.~R. Ellis, R.~Poltis, J.-P. Uzan, and A.~Weltman, ``{Blackness of the
  cosmic microwave background spectrum as a probe of the distance-duality
  relation},'' \href{http://dx.doi.org/10.1103/PhysRevD.87.103530}{{\em Phys.
  Rev.} {\bfseries D87} no.~10, (2013) 103530},
\href{http://arxiv.org/abs/1301.1312}{{\ttfamily arXiv:1301.1312
  [astro-ph.CO]}}.
%%CITATION = ARXIV:1301.1312;%%.

\bibitem{Bonvin:2016qxr}
C.~Bonvin, C.~Caprini, R.~Sturani, and N.~Tamanini, ``{Effect of matter
  structure on the gravitational waveform},''
  \href{http://dx.doi.org/10.1103/PhysRevD.95.044029}{{\em Phys. Rev.}
  {\bfseries D95} no.~4, (2017) 044029},
\href{http://arxiv.org/abs/1609.08093}{{\ttfamily arXiv:1609.08093
  [astro-ph.CO]}}.
%%CITATION = ARXIV:1609.08093;%%.

\bibitem{1992grle.bookS}
P.~{Schneider}, J.~{Ehlers}, and E.~E. {Falco}, {\em {Gravitational Lenses}}.
\newblock Springer-Verlag Berlin/Heidelberg/New York, 1992.

\bibitem{Bartelmann:1999yn}
M.~Bartelmann and P.~Schneider, ``{Weak gravitational lensing},''
  \href{http://dx.doi.org/10.1016/S0370-1573(00)00082-X}{{\em Phys. Rept.}
  {\bfseries 340} (2001) 291--472},
\href{http://arxiv.org/abs/astro-ph/9912508}{{\ttfamily arXiv:astro-ph/9912508
  [astro-ph]}}.
%%CITATION = ASTRO-PH/9912508;%%.

\bibitem{Fleury:2015rwa}
P.~Fleury, J.~Larena, and J.-P. Uzan, ``{The theory of stochastic cosmological
  lensing},'' \href{http://dx.doi.org/10.1088/1475-7516/2015/11/022}{{\em JCAP}
  {\bfseries 1511} no.~11, (2015) 022},
\href{http://arxiv.org/abs/1508.07903}{{\ttfamily arXiv:1508.07903 [gr-qc]}}.
%%CITATION = ARXIV:1508.07903;%%.

\bibitem{Bonvin:2011bg}
C.~Bonvin and R.~Durrer, ``{What galaxy surveys really measure},''
  \href{http://dx.doi.org/10.1103/PhysRevD.84.063505}{{\em Phys. Rev.}
  {\bfseries D84} (2011) 063505},
\href{http://arxiv.org/abs/1105.5280}{{\ttfamily arXiv:1105.5280
  [astro-ph.CO]}}.
%%CITATION = ARXIV:1105.5280;%%.

\bibitem{Challinor:2011bk}
A.~Challinor and A.~Lewis, ``{The linear power spectrum of observed source
  number counts},'' \href{http://dx.doi.org/10.1103/PhysRevD.84.043516}{{\em
  Phys. Rev.} {\bfseries D84} (2011) 043516},
\href{http://arxiv.org/abs/1105.5292}{{\ttfamily arXiv:1105.5292
  [astro-ph.CO]}}.
%%CITATION = ARXIV:1105.5292;%%.

\bibitem{Kaiser:1987qv}
N.~Kaiser, ``{Clustering in real space and in redshift space},''
{\em Mon. Not. Roy. Astron. Soc.} {\bfseries 227} (1987) 1--27.
%%CITATION = MNRAA,227,1;%%.

\bibitem{Matsubara:2004fr}
T.~Matsubara, ``{Correlation function in deep redshift space as a cosmological
  probe},'' \href{http://dx.doi.org/10.1086/424561}{{\em Astrophys. J.}
  {\bfseries 615} (2004) 573--585},
\href{http://arxiv.org/abs/astro-ph/0408349}{{\ttfamily arXiv:astro-ph/0408349
  [astro-ph]}}.
%%CITATION = ASTRO-PH/0408349;%%.

\bibitem{Montanari:2015rga}
F.~Montanari and R.~Durrer, ``{Measuring the lensing potential with tomographic
  galaxy number counts},''
  \href{http://dx.doi.org/10.1088/1475-7516/2015/10/070}{{\em JCAP} {\bfseries
  1510} no.~10, (2015) 070},
\href{http://arxiv.org/abs/1506.01369}{{\ttfamily arXiv:1506.01369
  [astro-ph.CO]}}.
%%CITATION = ARXIV:1506.01369;%%.

\bibitem{Matsubara:1999du}
T.~Matsubara, ``{The Correlation function in redshift space: General formula
  with wide angle effects and cosmological distortions},''
  \href{http://dx.doi.org/10.1086/308827}{{\em Astrophys. J.} {\bfseries 535}
  (2000) 1},
\href{http://arxiv.org/abs/astro-ph/9908056}{{\ttfamily arXiv:astro-ph/9908056
  [astro-ph]}}.
%%CITATION = ASTRO-PH/9908056;%%.

\bibitem{Reimberg:2015jma}
P.~H.~F. Reimberg, F.~Bernardeau, and C.~Pitrou, ``{Redshift-space distortions
  with wide angular separations},''
  \href{http://dx.doi.org/10.1088/1475-7516/2016/01/048}{{\em JCAP} {\bfseries
  1601} no.~01, (2016) 048},
\href{http://arxiv.org/abs/1506.06596}{{\ttfamily arXiv:1506.06596
  [astro-ph.CO]}}.
%%CITATION = ARXIV:1506.06596;%%.

\bibitem{Green:2014aga}
S.~R. Green and R.~M. Wald, ``{How well is our universe described by an FLRW
  model?},'' \href{http://dx.doi.org/10.1088/0264-9381/31/23/234003}{{\em
  Class. Quant. Grav.} {\bfseries 31} (2014) 234003},
\href{http://arxiv.org/abs/1407.8084}{{\ttfamily arXiv:1407.8084 [gr-qc]}}.
%%CITATION = ARXIV:1407.8084;%%.

\bibitem{Cusin:2016kqx}
G.~Cusin, C.~Pitrou, and J.-P. Uzan, ``{Are we living near the center of a
  local void?},'' \href{http://dx.doi.org/10.1088/1475-7516/2017/03/038}{{\em
  JCAP} {\bfseries 1703} no.~03, (2017) 038},
\href{http://arxiv.org/abs/1609.02061}{{\ttfamily arXiv:1609.02061
  [astro-ph.CO]}}.
%%CITATION = ARXIV:1609.02061;%%.

\bibitem{Fleury:2015hgz}
P.~Fleury, {\em {Light propagation in inhomogeneous and anisotropic
  cosmologies}}.
\newblock PhD thesis, Paris, Inst. Astrophys., 2015.
\newblock \href{http://arxiv.org/abs/1511.03702}{{\ttfamily arXiv:1511.03702
  [gr-qc]}}.
\newblock
\url{https://inspirehep.net/record/1404129/files/arXiv:1511.03702.pdf}.
\newblock
%%CITATION = ARXIV:1511.03702;%%.

\end{thebibliography}\endgroup

\end{document}